%% file: main.tex
\documentclass[preprint]{vgtc} 
\UseRawInputEncoding

\onlineid{0}
\vgtccategory{Research}

\input{util}

\begin{document}

\preprinttext{Published in: IEEE Transactions on Visualization and Computer Graphics 31,10 2025.\\ \url{https://doi.org/10.1109/TVCG.2025.3592962}}

\title{Augmented Reality Productivity In-the-Wild: \\
A Diary Study of Usage Patterns and Experiences of Working with AR Laptops in Real-World Settings}
\author{Yi Fei Cheng, Ari Carden, Hyunsung Cho, Catarina G. Fidalgo, Jonathan Wieland, and David Lindlbauer}
\authorfooter{
  \item Yi Fei Cheng, Ari Carden, Hyunsung Cho, and David Lindlbauer are with Carnegie Mellon University. 
  \item Catarina Fidalgo is with INESC-ID, Instituto Superior T\'ecnico, University of Lisbon and Carnegie Mellon University. 
  \item Jonathan Wieland is with the University of Konstanz. 
}

\keywords{Augmented Reality, In-the-wild, User Study}

\teaser{
  \centering
  \includegraphics[width=0.95\linewidth]{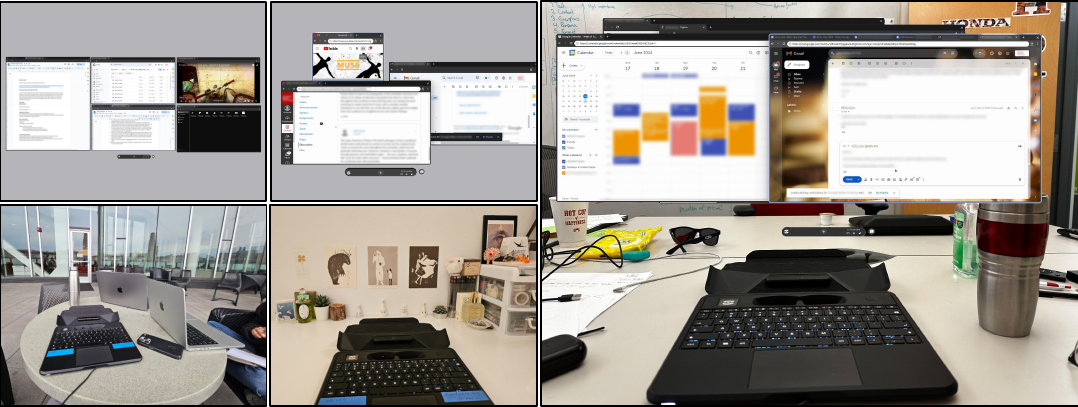}
    \caption{Our study explores in-the-wild usage of an Augmented Reality laptop. Fourteen participants configured personalized virtual workspaces to support their tasks within their daily productive environments, spanning 143 usage sessions. \mr{\emph{Left:} screenshots of participants' virtual workspaces (\emph{top}), captured from their devices, and the corresponding physical environments (\emph{bottom}), photographed by the participants. \emph{Right:} a synthesized approximation of a participant's viewpoint, composited from a device screenshot and the corresponding environment photograph.}}
    \label{fig:teaser}
}

\input{sections/abstract}

\maketitle

\input{sections/introduction}
\input{sections/related-work}
\input{sections/methods}

\input{sections/results}
\input{sections/discussion}

\input{sections/conclusion}
\input{sections/acknowledgements}

\bibliographystyle{abbrv-doi-hyperref}
\bibliography{references}

\input{sections/biography}

\input{sections/supplementary}

\end{document}

%% file: util.tex
\usepackage{xspace}
\usepackage{booktabs}
\usepackage{graphicx}

\newcommand{\eg}{e.g.,\xspace}
\newcommand{\ie}{i.e.,\xspace}
\newcommand{\etal}{et~al.\xspace}

\newcommand{\statsum}[3]{{$M$~=~#1,~$SD$~=~#2}}

\newcommand{\customtilde}{{\raise.17ex\hbox{$\scriptstyle\sim$}}}

\newcommand{\mr}[1]{#1}

%% file: sections/abstract.tex
\abstract{
Augmented Reality (AR) is increasingly positioned as a tool for knowledge work, providing beneficial affordances such as a virtually limitless display space that integrates digital information with the user's physical surroundings. However, for AR to supplant traditional screen-based devices in knowledge work, it must support prolonged usage across diverse contexts. Until now, few studies have explored the effects, opportunities, and challenges of working in AR outside a controlled laboratory setting and for an extended duration. This gap in research limits our understanding of how users may adapt its affordances to their daily workflows and what barriers hinder its adoption. In this paper, we present findings from a longitudinal diary study examining how participants incorporated an AR laptop --- Sightful's Spacetop EA --- into their daily work routines. 14 participants used the device for 40-minute daily sessions over two weeks, collectively completing 103 hours of AR-based work. Through survey responses, workspace photographs, and post-study interviews, we analyzed usage patterns, workspace configurations, and evolving user perceptions. Our findings reveal key factors influencing participants' usage of AR, including task demands, environmental constraints, social dynamics, and ergonomic considerations. We highlight how participants leveraged and configured AR's virtual display space, along with emergent hybrid workflows that involved physical screens and tasks. Based on our results, we discuss both overlaps with current literature and new considerations and challenges for the future design of AR systems for pervasive and productive use.
}

%% file: sections/introduction.tex
\section{Introduction}
Augmented Reality (AR) has the potential to change the way we engage in knowledge work.
Commercial off-the-shelf devices, such as the Apple Vision Pro~\cite{apple2023introducingvp} and Lenovo ThinkReality Glasses~\cite{lenovo2025thinkreality}, are increasingly promoted as productivity tools.
Previous research has also highlighted several benefits that AR offers in this context, such as reducing visual distractions and allowing the personalization of shared workspaces~\cite{lee2022partitioning}.
Furthermore, AR has been shown to support the in situ display of digital information~\cite{chen2023papertoplace} and facilitate collaborative embodied brainstorming and sensemaking~\cite{luo2022placement}.

A particularly promising affordance of AR for knowledge work is its ability to present information 
without being limited to physical monitors~\cite{pavanatto2024xrwild}.
Using a head-mounted display (HMD), 
application contents that were previously confined to computer screens can be seamlessly integrated into the user's physical environment, allowing them to be displayed anytime and anywhere, with variable size and appearance.
This support for an effectively infinite display space provides significant productivity benefits over conventional devices by minimizing context-switching costs and improving user awareness of peripheral information~\cite{mcgill2020seatedvrworkspace}.
These properties are especially useful in mobile scenarios, enabling work in public or confined spaces~\cite{biener2024xrworkpublic}.

Many companies and researchers suggest that AR will eventually replace computers and phones in the ``office of the future''~\cite{gruber2018officefuture}, enabling all-day knowledge work on the go. 
However, prior studies evaluating AR's support of knowledge work have predominantly been conducted in laboratory settings and over relatively short timeframes~(\eg~\cite{biener2024xrworkpublic,pavanatto2024xrwild}).
Although these earlier studies are valuable for validating the feasibility of performing knowledge work with current AR devices and informing future developments,
they may not fully reflect user experiences during everyday computing activities~\cite{abowd2000chartingubicomp}. 
Furthermore, the limited duration of these studies obscures our understanding of how this technology shapes people's behaviors once they have had time to adjust to its novelty~\cite{bailenson2024seeing}.
Therefore, we see it as beneficial to explore how people use AR for work in their daily environments over time.

In this paper, we present the results of a diary study investigating participants' usage of an AR laptop --- Sightful's Spacetop EA (Early Access) --- for daily work tasks and activities.
The Spacetop EA is an optical see-through AR HMD \mr{connected to a laptop base with} standard keyboard and trackpad for input~(Section~\ref{sec:apparatus}).
The device enables people to work on real productive tasks in their own physical environments.
Notably, it supports flexibly opening and arranging virtual windows for common applications used in knowledge work, similar to the interfaces explored by Pavanatto~\etal~\cite{pavanatto2024multiplemonitors}.
In our study, we deployed this AR laptop for 14 participants to use during their workday over the course of two weeks in 40-minute daily sessions.
During the study, we collected photographs of their virtual workspace and physical environment, along with post-session survey responses about their usage and perception of the device.
Participants collectively completed a total of 103 hours of work using the AR laptop (\customtilde7 hours per participant) across 143 sessions (\customtilde10 sessions per participant).
The study concluded with \customtilde30-minute interviews, encouraging participants to reflect more deeply on their overall experience.

Our analysis of the survey data, workspace photographs, and interviews provides insights into how considerations around task, environment, social, and physical comfort influenced participants' usage behaviors, including the number, size, and arrangement of windows within their virtual workspaces.
We also report on instances where participants naturally used the device to complement physical displays and tasks. 
Finally, our analysis highlights the current values and challenges of using an AR laptop for tasks in everyday work settings. 
In our discussion, we reflect on how our findings, derived from a more ecologically-valid and longitudinal approach to investigating AR usage, relate to prior empirical studies.
We also highlight implications for future design and research, including directions for developing interaction techniques, onboarding approaches, and adaptive systems. 

In summary, we contribute the following:
\begin{itemize}
    \item An analysis of AR usage patterns from a longitudinal diary study conducted in situ within people's daily work environments.
    We report on usage considerations, virtual display configuration patterns, and hybrid usage involving both physical displays and tasks
    \item Documentation of users' perceptions and experiences of AR's value for everyday tasks, as well as common challenges
    \item Implications and opportunities derived from our qualitative analysis, such as potential approaches to managing virtual workspaces, or how to better leverage contextual factors for assisting users 
\end{itemize}

As the AR research community and commercial stakeholders build toward a vision of ``pervasive'' AR usage~\cite{grubert2017pervasive}, our work aims to shed light on user behaviors and experiences that may arise from prolonged use in real-world settings, informing future technology designs to better accommodate these emerging needs and challenges.

%% file: sections/related-work.tex
\section{Related Work}
\label{sec:related-work}

\subsection{Productivity in Mixed Reality}
\label{sec:related-work-productivity}
In prior research, there is a significant body of work leveraging devices across the \mr{Mixed Reality (MR)} continuum, including Augmented and Virtual Reality (AR/VR), for productivity-related applications.
Many systems and prototypes have been developed to support tasks such as sensemaking~\cite{lisle2021immersivethink}, 
working with spreadsheets~\cite{gesslein2020spreadsheet},
and collaboration~\cite{pejsa2016room2room}.
These systems use a variety of interaction techniques, 
ranging from multi-device input for navigating information~\cite{le2021vxslate,biener2020breakingscreen} to gaze-based approaches for accessing applications~\cite{lu2020glanceable}. 

For many commercial stakeholders and researchers, MR represents a future pervasive medium for work, much like how computers and mobile phones are today~\cite{gruber2018officefuture,grubert2017pervasive,feiner2002ar}.
Numerous studies have highlighted the advantages that MR offers as a ubiquitous tool for work.
For instance, MR's ability to personalize a user's visual work environment 
has been shown to reduce distractions and stress while inducing a state of flow~\cite{ruvimova2020transportaway,lee2022partitioning,thoondee2017vrstress}. Schneider~\etal~\cite{reconviguration2019schneider} and Grubert~\etal~\cite{gruber2018officefuture} examined the potential privacy benefits of the HMD form factor, which restricts the visibility of content to the wearer.
Last but not least, MR relieves information displays of physical limitations~\cite{biener2022vrweek,gruber2018officefuture}. 
Using an HMD to present virtual content, including conventional windows, icons, and menus, MR provides more flexibility in where and how information is displayed~\cite{pavanatto2024xrwild}. 
As promoted by providers like Apple, Meta, and Sightful, it effectively provides a boundless workspace. 
This expanded screen real-estate may facilitate
complex multi-window operations by serving as a type of external memory~\cite{pavanatto2024multiplemonitors,andrews2010space,ball2007movetoimprove}, 
enable more efficient multitasking~\cite{czerwinski2003largedisplays},
provide quicker peripheral access to information~\cite{lu2020glanceable,davari2020occlusion,grudin2001digitalworlds}, 
and enhance immersion~\cite{endert2012lhrd}.

Yet, MR is still far from a commonplace tool for productivity due to significant usability challenges in current devices~\cite{ens2014personalcockpit,mcgill2020seatedvrworkspace,medeiros2022shieldingar,ng2021passengerexperiencemrairplane,pavanatto2024multiplemonitors,pavanatto2021virtualmonitor}. 
One longstanding issue is the resolution and field of view limitations of MR HMDs~\cite{pavanatto2021virtualmonitor,biener2024xrworkpublic}.
According to Pavanatto~\etal~ \cite{pavanatto2021virtualmonitor}, this reduces the effective screen space available to the user and increases head movement, which can cause neck pain \cite{gruber2018officefuture,mcgill2020seatedvrworkspace} and decreased task performance \cite{pavanatto2023virtualmonitor,pavanatto2021virtualmonitor}.
In AR, changes in focal distance when switching between virtual and physical content increase visual fatigue~\cite{gabbard2019arcontextswitch}
and interaction errors~\cite{eiberger2019depth}. 
AR also presents challenges in maintaining a balance between virtual and physical awareness. 
The visibility of virtual elements can sometimes be diminished by the physical environment~\cite{cheng2021semanticadapt}. 
On the other hand, interacting with virtual elements may distract users and reduce their awareness of their surroundings~\cite{tao2022distractions,li2022mrworkspaces}, including the presence of bystanders~\cite{ohagen2020realityaware}.
Finally, 
the increased screen real estate that MR offers may present challenges, such as difficulties in managing multiple windows and accessing distant information~\cite{endert2012lhrd,pavanatto2024multiplemonitors,czerwinski2003largedisplays}.

Overall, the literature currently describes a range of positive and negative aspects of working in MR. However, these insights were often derived from short-term studies conducted in laboratory environments. 
In contrast, our work aimed to examine the potential values and challenges of productive AR usage in a more ecologically valid context.

\subsection{Window-based Mixed Reality}
There has also been a persistent interest in using MR devices to enhance knowledge work by expanding traditional 2D displays and windows beyond the limitations of physical screens and into the third dimension~\cite{pavanatto2024multiplemonitors}.
Early work, such as Feiner~\etal~\cite{feiner1993windows}, 
explored various methods to register windows,
including head-fixed and world-fixed approaches, 
as well as how to create connections between windows and physical objects.
Raskar~\etal~\cite{raskar1998officefuture} investigated the possibilities of expanding an office space with AR using projectors and cameras.
Many recent commercial MR workspaces, such as the Varjo Workspace~\cite{varjoworkspace}, 
Immersed~\cite{immersed},
and vSpatial~\cite{vspatial},
now enable streaming of desktop elements into MR as virtual windows and support conventional input devices (\ie~mouse and keyboard) for interaction.
Using conventional 2D displays and desktop input devices for MR can ease the transition to virtual displays for users coming from personal computers, while also providing access to familiar and highly refined interfaces and applications~\cite{pavanatto2024multiplemonitors}.
Therefore, in our study, we focus on emergent user behaviors when using an AR laptop that features a window-based immersive display along with keyboard and touchpad input.

Several prior studies have investigated the usage of window-based MR. 
For example, Ens~\etal~\cite{ens2014ethereal} classified window-based mixed reality workspaces in seven dimensions, such as whether the windows used an ego- or exocentric reference frame or were movable.
Several prior studies~\cite{cheng2021semanticadapt,pavanatto2024xrwild,lischke2016screenarrangement} observed that users preferred to center main task content while using peripheral areas for supporting information.
Su~and~Bailey~\cite{su2005positioning} recommended against placing virtual contents behind the user.
McGill~\etal~\cite{mcgill2020seatedvrworkspace} reported on several common window configurations for general productivity tasks, including vertical, horizontal, two-plus-two, and following a ``personal cockpit''~\cite{ens2014personalcockpit}.
As immersive technologies become more pervasively used, 
the context of use also becomes a critical consideration~\cite{grubert2017pervasive}.
For example, the physical environment can hinder or enhance interactions~\cite{cheng2023interactionadapt}.
Similarly, social acceptability has been shown to play an important role in the use of virtual monitors in airplanes~\cite{ng2021passengerexperiencemrairplane} and other shared transit modalities~\cite{medeiros2022shieldingar}.
While many usage patterns have been observed in shorter laboratory studies, our research investigates how these patterns may manifest in more realistic environments and over extended periods.

\subsection{In-the-Wild and Longitudinal Studies of MR}
There is a growing number of in-the-wild and extended-use studies of MR systems.
In-the-wild studies of MR have demonstrated effectiveness across various application areas, including education~\cite{petersen2021pedagogical},
accessibility~\cite{schmelter2023accessible},
military training~\cite{laviola2015military}, 
and \mr{on-the-go information access~\cite{chang2024walking,lu2021glanceableauthentic}.}
Researchers have also investigated the long-term effects of MR usage on depth perception~\cite{kohm2022objects}, social interactions~\cite{han2023longitudinal,bailenson2006longitudinal}, and within industrial scenarios~\cite{grubert2010mobileindustrialar}.
The most relevant to our work are previous studies on the use of MR for everyday productivity tasks. 

In early investigations, many researchers engaged in informal self-experimentation. 
Starner~\cite{stevens2013starner}
and Mann~\cite{buchanan2013mann}
independently designed and wore AR headsets almost daily for over a decade.
Recent studies have continued this line of work with more participants, often with the aim of quantifying the effects.
For example, Steinicke and Bruder~\cite{steinicke2014selfexperimentation}
and Nordahl~\etal~\cite{nordahl2019hours} examined the impact of prolonged VR usage on key usability factors, including simulator sickness, over periods of 24 hours and 12 hours, respectively.
Lu~\etal~\cite{lu2023inthewild} evaluated a glanceable AR prototype with three participants over three days.
Guo~\etal~\cite{guo2019maslows}
and Shen~\etal~\cite{shen2019longtermfatigue} 
report on a study in which 27 participants worked eight hours each in virtual and physical office environments, focusing on emotional and physiological needs, as well as visual and physical discomfort.
Biener~\etal~\cite{biener2022vrweek,biener2024holdtight} compared VR work with a regular physical environment over five days with 16 participants. 
Pavanatto~\etal~\cite{pavanatto2024xrwild} recently examined user and bystander experiences of Extended Reality~(XR) usage in public, studying how users perform their own tasks in XR at various campus locations (\eg~library, dining).
\mr{
Complementing these empirical studies, Tran~\etal~\cite{tran2025everdayyoutube} observed a keen interest in using MR for productivity through an ethnography of YouTube videos.
}

Both in-the-wild and longitudinal studies are valuable for examining user behaviors and attitudes in more ecologically valid settings~\cite{abowd2000chartingubicomp} and to bypass distortions from novelty effects~\cite{bailenson2024seeing}.
We build on existing work by reporting on a study in which we deployed an AR laptop for daily use in participants' everyday work environments over the course of two weeks. 
As a primary point of differentiation from most previous research,
we did not constrain the environment or time in which participants performed their own tasks in AR. 
Our device also allowed participants to arrange multiple virtual windows flexibly. 
Overall, we aimed to study a closer approximation of AR usage for everyday productivity tasks in the future, observing how individuals may appropriate the device to suit their specific needs and usage contexts~\cite{rogers2011wild}.

%% file: sections/methods.tex
\section{Study}
In this study, our goal was to gain insight into how knowledge workers may use an AR device for their daily activities and tasks.
In particular, we focus on the user behaviors and experiences of working with an AR laptop, 
which uses desktop input devices (\ie~keyboard and trackpad) to allow interaction with traditional 2D windows and applications rendered on an HMD.
In contrast to current laptops, our device provides an expanded virtual display space. 
We were particularly interested in three research questions related to AR laptop usage:
(1)~What are the core considerations affecting its use?
(2)~How do these considerations influence participants' configuration and usage of their AR workspace, such as the number and arrangement of virtual windows?
(3)~What are the values and challenges of using an AR laptop?


\subsection{Apparatus}
\label{sec:apparatus}

Our apparatus 
was informed by the following considerations.
First, the device should have a design and support input methods that enable prolonged interaction~\cite{cheng2022comfortable}.
This suggests that the HMD should ideally be lightweight for user comfort. 
Furthermore, since keyboards and trackpads are generally more efficient and ergonomic than other XR input methods such as speech or gestures for knowledge work-oriented tasks~\cite{zhou2022indepthmouse,knierim2018physicalkeyboardsvr}, our device should incorporate these traditional input modalities.
Second, the device should support common applications for knowledge work, such as an email client, word processor, and web browser. 
In line with Pavanatto~\etal~\cite{pavanatto2024multiplemonitors}, we assume that in the near future, AR will be designed around 2D interfaces to ease the transition from today's personal
computers to computing with virtual displays. 
The device should use familiar interface designs while incorporating AR functionalities for an expanded in situ display space.
 
\begin{figure}[t]
    \centering
    \includegraphics[width=\columnwidth]{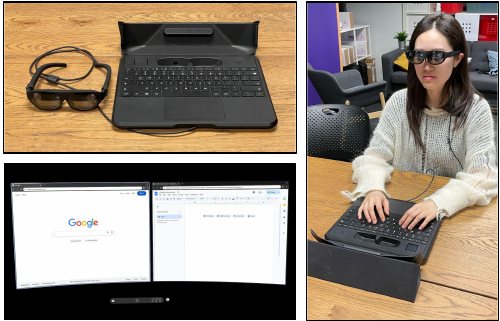}
    \caption{Spacetop EA device (\textit{top left}) and interface (\textit{bottom left}), and a user wearing the device (\textit{right}). Note that for users the background appeared transparent (\ie optical see-through), not black as depicted.}
    \label{fig:device}
    \vspace{-1em}
\end{figure}

After pilot testing several devices and approaches, we chose to use the Sightful Spacetop EA (Early Access), shown in Figure~\ref{fig:device}.
This \mr{standalone} device consists of a pair of optical see-through display glasses (Xreal Light, $85$ grams, OLED display panels, $1920 \times 1080$ pixels per eye, $72$ Hz refresh rate, $~52^\circ$ diagonal field of view) for output attached to a standard keyboard and trackpad for input.
It has a Qualcomm Snapdragon XR2 processor, 
a Kryo 585TM 8-core 64-bit CPU, an AdrenoTM 650 GPU, and 8GB of LPDDR5 RAM.
The device runs a custom operating system, Spacetop OS, which enables users to flexibly open, close, resize, and move one or more window-based web applications (\eg~Google workspace application, WhatsApp, ChatGPT) on top of a virtual ``spatial canvas'' --- a unified display space that is curved over a cylinder surrounding the user, similar to the concept described by Pavanatto~\etal~\cite{pavanatto2024multiplemonitors}.
The spatial canvas is notably world-anchored~\cite{feiner1993windows} and can be zoomed in and out of, panned from side to side, height adjusted, tilted, and re-centered over the device keyboard.
\mr{Participants could adjust the distance and size of windows and fonts to be readable comfortably. We qualitatively confirmed that at a 2-meter distance between the user and the virtual windows, a font size of 12 pt could be read comfortably, with visible degradation of legibility for fonts smaller than 8 pt. Legibility was sufficient for everyday productivity tasks, \eg~working in browser windows.}
\mr{The device does not have a front-facing camera, and does not have the capability to track the environment or other devices for spatial anchoring.}
\mr{
The device's technical specifications and supported interactions are described in detail in Supplementary Material.
We did not make study-specific customizations to the device software for our deployment.
}
We chose the device over video passthrough options like the Meta Quest 3~\cite{metaquest3} or Apple Vision Pro~\cite{apple2023introducingvp} mainly because of its lighter and more ergonomic glasses form factor, which is essential for prolonged productive tasks~\cite{biener2022vrweek}.

\subsection{Experimental Design and Procedure}
To address our research questions, 
we designed a longitudinal, in situ diary and interview study aimed at understanding people's usage patterns and perceptions.
As discussed by Bailenson~\etal~\cite{bailenson2024seeing},
longitudinal studies with AR are critical to understanding behaviors and challenges associated with its anticipated ``all-day use.''
Abowd~and~Mynatt~\cite{abowd2000chartingubicomp} highlighted the need for evaluations to be carried out in everyday computing environments, rather than solely in controlled laboratory settings, to achieve greater ecological validity.

Our particular approach was guided by several studies of methodological similarity (\eg~\cite{czerwinski2004diary,borghouts2022diary,jokela2015diary}).
Diary studies are a common qualitative research method in HCI~\cite{consolvo2017diary}, widely used to understand behavior over longer periods, such as working within multi-device ecologies~\cite{jokela2015diary} or work-from-home conditions during the COVID-19 pandemic~\cite{borghouts2022diary}.
The study was approved by the \mr{Institutional Review Board (IRB)} of the participating research institutions.

Our study was conducted in three phases: an initial onboarding session, a two-week usage period, and a final interview.
All study protocols and surveys are documented in the Supplementary Material. 

\paragraph{Onboarding.}
Recruited participants were invited into the laboratory for an hour-long onboarding session. 
During this session, they were first informed about the procedure and content of the study. 
Following this, they signed a consent form and completed a background questionnaire. 
Afterward, the participants were introduced to the AR laptop and its key features. 
The experimenter guided them through the operations of turning the device on and off, adjusting the display brightness, opening and modifying windows, and moving and recentering their virtual canvas.
Finally, participants were given time to freely explore the features of the device and ask questions accordingly.
After participants indicated their comfort with the device controls, they were given the device to use in their own work environments during the second phase of the study. 
We also shared a reference guide to navigate the core functionalities of the device and encouraged participants to reach out if they encountered any issues or questions.

\paragraph{Usage and Diary Study.} 
We deployed the device to each participant for two weeks (14 days).
\mr{During the usage period, participants were instructed to complete at least one 30-minute session on 10 of the 14 study days.}
Participants \mr{could complete as many additional sessions as they wished} for extra compensation.
\mr{This means they used the device during at least 10 different days.}

Participants were encouraged to use the device during their daily activities and tasks, which they would typically complete regardless of their participation in the study, in the familiar locations they frequented as part of their routines.
We deliberately chose not to set specific constraints on the task and environment, in order to gather more ecologically valid data on participants' usage behaviors.
We set individual sessions at a minimum of 30 minutes to ensure that participants had enough time to engage in meaningful tasks.
We evaluated usage across 10 sessions to allow participants to acclimate to the device and understand how their usage changes with different tasks and extended exposure, without overburdening them. 

After each session, participants were tasked with completing a diary entry, administered as a survey. 
The survey prompted participants to report on
what tasks they completed on and off the device, 
where they completed the task, 
and when the session started and ended.
It also asked how participants organized their virtual workspaces, any adjustments they made, and the factors that influenced these processes. 
Moreover, it inquired about participants' likes and dislikes regarding the device.
Finally, it asked participants to optionally upload photographs of their physical surroundings and virtual workspace.
The survey took around 15 minutes to complete. 
Participants received daily emails reminding them to complete their session and diary entry. 
We encouraged them to fill out the diary entry immediately after their sessions. 

We conducted pilot tests of the survey with four knowledge workers, who were recruited through snowball sampling. 
\mr{
1 pilot participant completed 2 sessions (0.9 hours of total usage time), 2 completed six each (3.7 and 5.0 hours), and 1 completed 7 (4.9 hours). 
}
These pilot sessions helped us identify core questions, clarify wording, and ensure data quality. 
We revised the survey after each round of feedback. 
The results from these pilot sessions were not included in our analysis.

\paragraph{Interview.}
At the end of each participant's usage sessions, 
we reviewed their diary entries in advance to prepare for the semi-structured interviews.
We organized the interview into four blocks.
We first asked about the tasks performed on the device, whether any tasks were notably well or poorly suited, and how their experience using AR for these tasks compared to conventional devices.
Second, we asked how participants organized their virtual workspaces, how these workspaces evolved over time and across sessions, and the factors they considered.
We then investigated how participants engaged in non-AR activities (\eg~checking their phones) while using the device.
At the end of the interview, participants were asked to summarize the main benefits and drawbacks of the device, and to envision how AR might be used in knowledge work in the future.
Throughout the interview, participants could refer back to their diary entries, which we collated into a spreadsheet.
We also often asked participants to provide more details about some of their responses.
Interviews lasted around 30 minutes and were audio recorded. 
Half of the participants (N=7) completed the interview in-person, while the other half participated remotely.

\paragraph{}
Participants received $\$150$
for completing ten usage sessions and the exit interview.
They received $\$15$ for each additional session they completed beyond the minimum requirements.
Usage sessions took place from March 19, 2024, to August 29, 2024.
We deployed between two and three devices simultaneously to speed up data collection.

\subsection{Participants}
We recruited 14 people (aged 22 to 56 years, \statsum{30}{11}{}; 9 male, 5 female) with normal or corrected vision to participate in the study. 
We recruited university students, researchers, and staff for our study due to their diverse tasks, extensive computer use in their work, and availability. 
Participants self-reported low to moderate experience with AR (\statsum{2.6}{1.8}{}) and VR (\statsum{3.1}{1.7}{}) on a scale from 1 (low) to 7 (high).
Participants indicated that they regularly use 2 to 7 (\statsum{4}{1}{}) digital devices (\ie~mobile phones, laptops, tablets, and smartwatches).
They reported spending approximately 5 to 14 hours each day (\statsum{7}{2}{}) on their primary device, which was either a laptop or desktop computer.
All participants reported working remotely to some extent, typically at least once a week.
They also reported having normal or corrected-to-normal vision.

\subsection{Data Collection and Analysis}
\label{sec:analysis}
We collected our data from three main sources:
written diary survey entries,
workspace photographs, 
and interview audio recordings.
Participants collectively submitted a total of 143 diary entries. 
Ten participants completed 10 entries each, two participants completed 11 entries, and one participant completed 12 entries. 
One participant submitted one fewer entry than required (9 entries) due to an experimenter's miscalculation.
On average, each session lasted $42$ minutes ($SD$ = 16). 
In total, the diary entries documented 103 cumulative hours of work (\statsum{7.3}{1.6}{} hours per person). 

In addition, participants collectively submitted 140 photographs of their virtual workspaces and 109 photographs of their physical environments. 
All interviews with participants were 
\mr{
automatically transcribed and then reviewed and corrected by the first author.
}

\begin{figure*}
    \centering
    \includegraphics[width=0.9\linewidth]{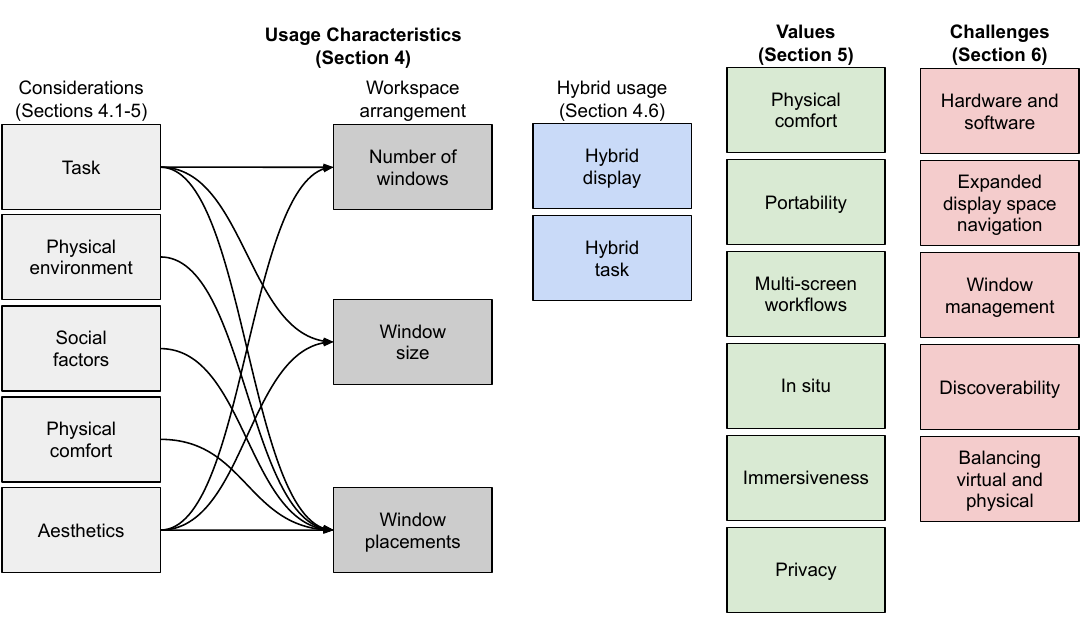}
    \caption{\mr{Overview of the \textbf{themes} that surfaced during our analysis}. We report on how considerations, such as the participants' tasks, influenced their workspace arrangement behaviors, such as the number of windows they opened. In addition, we highlight alternative workflows that involve the use of AR in tandem with physical displays and tasks. Finally, we summarize participants' perceptions of the value and challenges of using AR.}
    \label{fig:themes}
\end{figure*}

For our analysis, we adopted a similar methodology as Biener~\etal~\cite{biener2024holdtight} and Yuan~\etal~\cite{yuan2022multideviceusage}, using \emph{a priori} and open coding to create a codebook to characterize participants' usage and perceptions of the AR laptop.
We started our codebook based on our research questions and related literature (see Section~\ref{sec:related-work}), which documented various AR usage patterns (\eg~window arrangements), values, and challenges.
We then iteratively refined and expanded our codes using the collected data.
In each round of coding, 
two authors independently reviewed and labeled all survey, image, and interview data with existing codes, while defining new codes for relevant behaviors and comments that had not been previously captured.
The authors then convened to discuss the resulting set of codes and their respective scopes, merging similar codes and consolidating them into a shared codebook.
An affinity diagram was often used to help facilitate this process.

The final codebook comprised 171 individual codes \mr{(see Supplementary Material)} that detailed the tasks participants performed using the AR laptop, 
the physical environments in which they worked,
how they organized their virtual workspace for their tasks,  
and their perceptions regarding the value and limitations of the device.
Using this shared codebook, 
two authors collaboratively conducted a final round of coding on the original data. 
They applied the codes from the codebook to each instance, labeling every survey entry and interview with zero or more codes, based on a unanimous decision.
We do not report inter-rater reliability because following best practices from Hammer and
Berland~\cite{hammer2014irrcritique}, coding for each instance was unanimously agreed upon by annotators.
The same two authors also led the clustering process, organizing the codes into themes. 
These themes were refined iteratively with feedback from the remaining members of the research team, resulting in the findings presented below.

%% file: sections/results.tex
\section{Usage Characteristics}
In this section, we present our results on how participants used AR laptops during the two-week diary period. Several factors influenced their device usage, including 
\emph{task}, 
\emph{physical environment}, \emph{social interactions}, 
\emph{physical comfort}, 
and \emph{aesthetics}. 
In addition, we report on \emph{hybrid use cases} in which the AR laptop was used simultaneously with physical devices and tasks. 
We present the frequency of behaviors and remarks based on our closed coding results~(Section~\ref{sec:analysis}) to reflect the importance of a situation rather than an accurate measurement of how often a situation occurs~\cite{myers2016programmers}.
We use $n$ to denote the number of sessions during which a code was observed, out of 143 total sessions across all participants~(\eg~participants collectively watched videos in $n$~=~59 sessions).
We use $N$ to denote the number of participants associated with each code~(out of 14; \eg~$N$~=~10 participants watched videos at some point throughout the study).
\autoref{fig:themes} summarizes the main themes that we report on in our results.

\begin{figure}[t]
    \centering
    \includegraphics[width=.95\columnwidth]{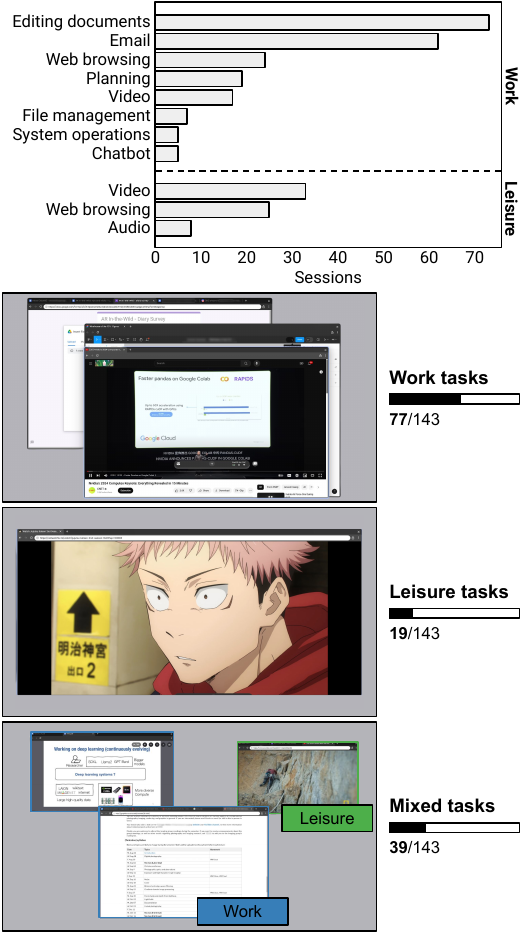}
    \caption{\textbf{Tasks} --- Participants engaged in a variety of work and leisure tasks. In most sessions, they focused on productive work but also explored leisure activities and occasionally mixed work with leisure.}
    \label{fig:tasks}
\end{figure}

\begin{figure}[t]
    \centering
    \includegraphics[width=0.95\linewidth]{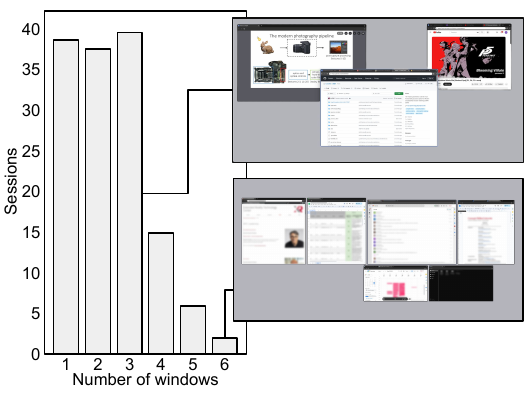}
    \caption{\textbf{Number of windows} --- Participants used an average of two windows, with a maximum of six. Most sessions involved three or fewer.}
    \label{fig:windows}
\end{figure}

\subsection{Task}
Participants reported performing a variety of tasks with the AR laptop during their usage sessions (\autoref{fig:tasks}).
The most common activities included reading or editing documents and other artifacts~($n$~=~74,~$N$~=~14), 
followed by checking and composing emails~($n$~=~63,~$N$~=~10),
watching videos~($n$~=~59,~$N$~=~13), 
and browsing web content~($n$~=~54,~$N$~=~14).

During a single session, participants often \textbf{engaged in activities across multiple task categories}~($M$~=~2,~$SD$~=~1~tasks~per~session). 
By categorizing activities as either productivity- or leisure-oriented, we found that participants primarily focused on productivity-oriented tasks~($n$~=~77,~$N$~=~14).
Some participants also explored more leisure-oriented tasks~($n$~=~19,~$N$~=~7)
or combined productive and leisure activities within the same session~($n$~=~39,~$N$~=~10).

Participants frequently reported that their tasks influenced how they configured their AR workspace~($n$~=~116,~$N$~=~13).

First, \textbf{their tasks influenced the number of virtual windows they opened} (\autoref{fig:windows}).
Our analysis showed that our participants configured virtual workspaces containing an average of 2.4 windows ($SD$~=~1.2).
We observed a maximum of six windows used simultaneously~($n$~=~2,~$N$~=~2);
however, workspaces containing four or more windows~($n$~=~23,~$N$~=~8)
were notably less common than those with three or fewer~($n$~=~118,~$N$~=~14).

\begin{figure*}
    \centering
    \includegraphics[width=0.95\textwidth]{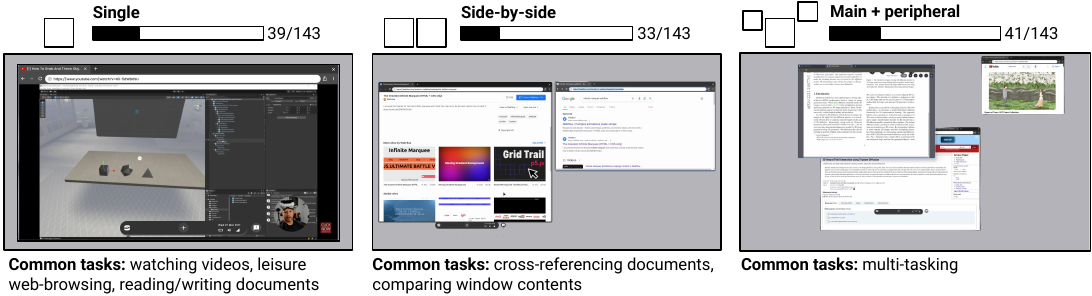}
    \caption{\textbf{Arrangement patterns} --- Common window arrangement patterns include using a single window, placing windows side-by-side, and positioning secondary windows in the periphery.}
    \label{fig:patterns}
\end{figure*}

In 39 sessions, participants configured their AR workspace to consist of only a single window (\autoref{fig:patterns},~left).
This layout was most frequently used for watching video~($n$~=~14,~$N$~=~7), browsing websites~($n$~=~10,~$N$~=~5), and reading and writing documents~($n$~=~7,~$N$~=~3).
For productive use cases, 
several participants~($N$~=~5) reported that this layout helped them focus on a single task.
As P5 commented, ``I did not feel the need to open more windows.... I would often open multiple windows when I am multitasking.''
P11 used a single window layout to read because it was ``free from distractions''.
For entertainment purposes, like watching a video, participants appreciated the immersiveness of this arrangement~($N$~=~4).
As P10 remarked, ``I placed [a single] window in the middle to have a sense of a big TV or cinema''.

In 33 sessions, 
participants placed two windows side by side~(\autoref{fig:patterns},~middle). 
Participants reported that \textbf{arranging windows side-by-side facilitated cross-referencing and comparison tasks}.
For example, as P8 recalled, 
``scheduling was nice because I could write an email and glance into my schedule without having to go between tabs.''
P7 shared a similar experience: ``I was watching a video on prompt engineering, and then on the right, I was using ChatGPT to try to ask certain questions a certain way versus different ways.... It was really useful for having that extra screen, just to be there and learn from.''

In 66 sessions, 
participants configured their AR workspaces to include three or more windows.
\textbf{Participants generally leveraged these multi-window workspaces to facilitate multitasking}.
P1, for instance, appreciated ``the way I can do multitasks, where I have some displays used for entertainment next to displays that are used for work stuff.''
As with many other participants~($n$~=~32,~$N$~=~11), 
they often worked on one or two main windows while playing a video or listening to music in other windows (\eg~\autoref{fig:patterns},~right).
Similarly, P5 reported that within one workspace, they would maintain windows for their primary task alongside ones they were ``not actively using.''

Generally, \textbf{participants organized their windows based on their expected usage frequency}, positioning windows for \textbf{primary tasks at the center} and moving \textbf{secondary windows to the periphery} or outside their field of view~($n$~=~41,~$N$~=~11;~\autoref{fig:patterns},~right).
For example, P4 described their workflow: ``I kept my primary working screen immediately in front of me (typically reserved for email). Supporting screens, such as Google Drive or Google Sheet, were placed to the left. My calendar remained below the main window.''
Similarly, P13 mentioned, ``I push non-critical things to the corners of my field of view, so that they are easily accessed, but not interfering with my work.'' 

Several participants ($N$~=~4)
also \textbf{used their peripheral screen space in AR to host persistent windows across multiple sessions to support ongoing or future tasks}. 
P4 reported moving windows ``way off to the side,'' then ``mov[ing] them back in'' when needed. 
This included a persistent email window that they stored ``in the corner somewhere, that I could glance at.''
For several sessions, P14 used one window to accumulate readings and resources they anticipated potentially returning to for their work.

\begin{figure*}
    \centering
    \includegraphics[width=0.95\textwidth]{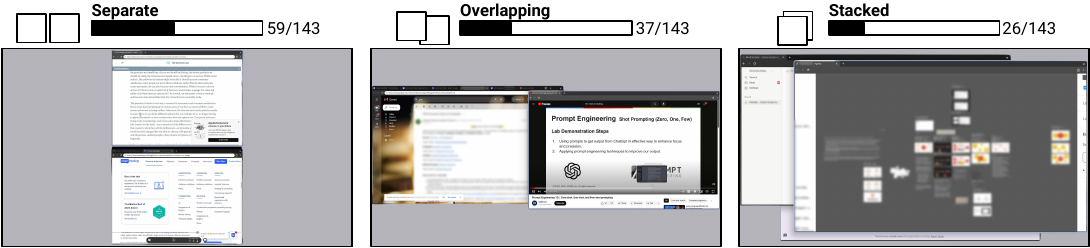}
    \caption{\textbf{Window spacing} --- The majority of participants preferred to keep windows separate, while a few opted to overlap or stack them.}
    \label{fig:spacing}
\end{figure*}

Participants mainly arranged windows in a way that separated them in space~($n$~=~59,~$N$~=~11;~\autoref{fig:spacing},~left), 
ensuring that their contents were fully visible. 
For P8, \textbf{separating windows facilitated their tasks by allowing them to ``see them all at a glance}, without having them stacked on top of each other like a conventional laptop does.''
Participants also overlapped~($n$~=~37,~$N$~=~11;~\autoref{fig:spacing},~middle) or stacked~($n$~=~26,~$N$~=~11;~\autoref{fig:spacing},~right) windows.
P3 and P10 reported that \textbf{stacking windows facilitated efficient task switching by allowing them to quickly ``swap back and forth'' between them}.

Besides the number of windows and their arrangement, 
\textbf{participants' tasks also influenced how they sized windows in AR}.
Several participants reported expanding single windows to form factors unsupported by conventional displays~($n$~=~17,~$N$~=~8).
For example, 
P6 found it beneficial to extend windows lengthwise to ``see more content at a time without scrolling.''
They specifically used longer windows to fill out forms and examine designs and documents in full, providing a more holistic overview.
P11 noted that larger windows were ``especially good for editing and reading.'' 
13 participants 
used enlarged windows to watch videos. 
P5 commented, ``I like the fact that I can make my own theatre at home. The screen can be as big as I need it to be.''

Finally, during sessions, \textbf{several participants frequently adjusted their windows} --- including their size~($n$~=~38,~$N$~=~11), 
depth~($n$~=~37,~$N$~=~9), 
and placement~($n$~=~36,~$N$~=~11) --- 
partly \textbf{in response to task needs}~($N$~=~37,~$n$~=~11).
P11 reported alternating between ``bring[ing] things forward'' to ``read in depth'' and ``push[ing] them back'' to ``scan.''
P2 recalled re-organizing their workspace and bringing tasks to a ``more primary position'' as their focus shifted.

\subsection{\mr{Physical} Environment}
\label{sec:environment}
\begin{figure*}
    \centering
    \includegraphics[width=0.95\textwidth]{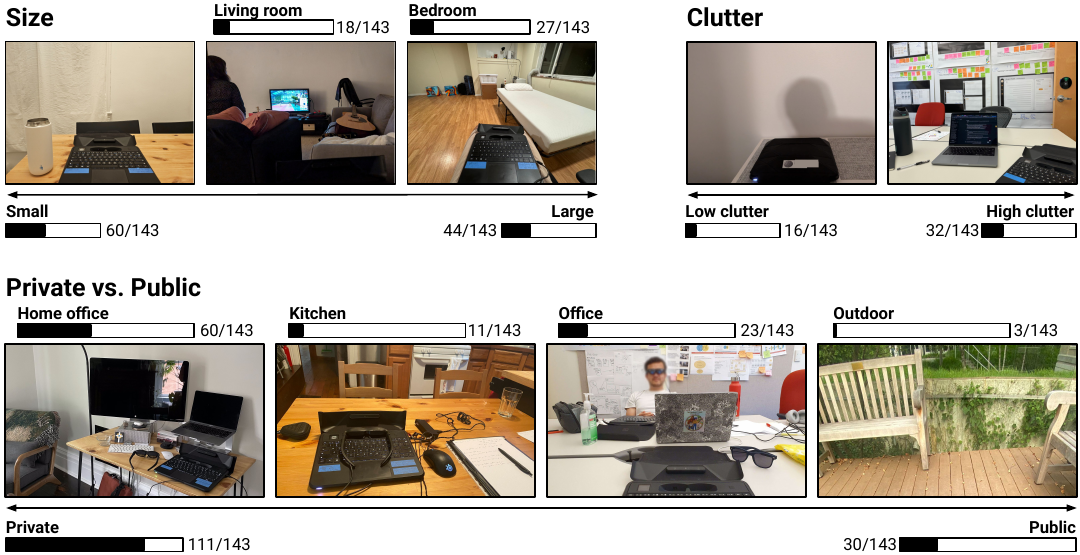}
    \caption{\textbf{Environments} --- Participants completed their sessions in a variety of physical environments. Their physical surroundings differed in size and level of visual clutter. Participants mostly worked in private spaces.}
    \label{fig:env}
    \vspace{-1em}
\end{figure*}

Participants completed their sessions in a variety of environments (\autoref{fig:env}), including
their home office~($n$~=~60,~$N$~=~9), 
bedroom~($n$~=~27,~$N$~=~9), 
office~($n$~=~23,~$N$~=~10),
living room~($n$~=~18,~$N$~=~5), 
and kitchen~($n$~=~11,~$N$~=~4).
The settings in which the participants worked varied in size (\autoref{fig:env},~top-left). 
Some used compact desk surfaces placed against a wall~($N$~=~60,~$n$~=~11), 
while others were in spacious areas that offered ample room in front of their devices~($n$~=~44,~$N$~=~10).
In addition, the environments differed in terms of visual clutter (\autoref{fig:env},~top-right). 
Some participants worked in clean spaces with plain walls~($n$~=~16,~$N$~=~5) 
while others were in heavily decorated environments~($n$~=~32,~$N$~=~10). 
Most sessions were completed against physical backgrounds that included a moderate number of objects~($n$~=~74,~$N$~=~12).

The AR laptop used by our participants overlaid virtual elements onto their physical surroundings. 
This made participants' usage behaviors context-sensitive to their environment~($n$~=~8,~$N$~=~13).

First, many \textbf{participants faced challenges with visibility of their virtual windows due to lighting conditions}~($n$~=~16,~$N$~=~7) \textbf{and background clutter}~($n$~=~18,~$N$~=~8)\textbf{, 
avoiding both in their window arrangements}.
P10, for instance, recalled having difficulties using the AR laptop in a bright office environment: ``It didn't work because there was so much light coming in, so what I did was push the windows to the ceiling.''
To minimize distractions caused by visual clutter, 
eight participants
reported that they intentionally sought out simpler walls to project their virtual windows.  
As P4 remarked, ``I appreciated just having a blank wall, and it was a neutral color, like an off-white, grayish.... So I thought that was a good environment.''

Here, it is important to note that the effects of environmental lighting and visual clutter may have been partly due to the display limitations of the device, particularly the rendering capabilities of its optical pass-through display. 
External lighting reduced the display contrast. 
As P12 commented, ``If I'm outside with the conditions and the brightness and the sun, I cannot be writing a document and reading a lot because the contrast is not enough for me.'' 
Reflecting on their experience of working in an office with sticky notes posted around the walls, 
P7 noted, 
``If there was a way to, like, completely make it opaque, where the virtual screen was completely opaque, ... it would be very helpful. I wouldn't have to see the sticky notes in the background very much [and] just be able to focus.''
This consideration also likely contributed to the participants' preference for completing their study sessions in indoor environments. 
Only three participants attempted working outdoors. 
As P9 remarked, ``I tried to sit outside like around lunchtime one day and that was a disaster because of the light.'' 

Interestingly, instead of avoiding visual clutter, two participants
used their virtual windows to shield themselves from distractions in their physical surroundings, creating a more focused workspace.
P14, for instance, 
reported placing a window over a TV in the background in one session and over a fan in another.

Besides visibility concerns, 
\textbf{participants' configurations of their virtual workspace were also generally guided by physical restrictions}.
Participants generally avoided overlapping virtual elements with physical objects and boundaries~($n$~=~19,~$n$~=~9).
For instance, P1 reported, ``I wanted the displays to be above my physical objects on my desk so that it doesn't interfere.''
Three participants
adjusted their workspaces according to how spacious they perceived their surroundings to be. 
In ``a very spacious room with high ceilings,'' P12 ``took advantage of that space'' to construct a wider AR workspace with three windows, whereas when the room ``was not very spacious,'' they only used a single virtual window.
However, several participants ($N$~=~3)
sometimes chose to ignore physical constraints. 
P1 recalled, ``since I was working on a personal desk with a white wall close in front of me, I placed the displays further away from me virtually to feel more like I am in a spacious room.'' 

One participant (P4) reported \textbf{arranging windows based on their semantic relationship with the physical environment}. 
In several sessions, they placed a weather application next to their window so that whenever they glanced over, they were informed of the outside temperature. 
P4 commented, ``I thought it was convenient to have it so closely associated with what I was looking at.''

\subsection{Social Factors}
\label{sec:social}
In 30 sessions, 
participants ($N$~=~8) used the AR laptop in public spaces or during interactions with others (\autoref{fig:env},~bottom).

Three participants
found it \textbf{comfortable to use the device in public}.
Five participants 
highlighted its \textbf{potential for keeping their work private while working in such spaces}. 
As P4 commented, ``The biggest benefit I saw was that if you're on an airplane and you want total secrecy, you don't want people looking over to see what you're typing in your email or the kind of work you're doing.'' 
P2 noted that the device provided significant \textbf{value during in-person meetings}. 
They reported that it allowed them to be ``immersed'' in their information environment while still staying ``a part of the team.''

On the other hand, five participants
voiced \textbf{concerns about the device's current social acceptability}.
As P1 noted, ``I actually wanted to use it outside... but my friends all told me that I'd look really weird and freaky.''
P9 recalled, ``One night I was doing this study and my husband came in. He was like, `what the hell are you doing?' It kind of freaked him out a little bit... then I started thinking, this does look weird. I don't know if I would feel comfortable to walk into Starbucks or sit on my front porch with all my nosy neighbors walking by.'' 
These issues with social acceptability were attributed partly to the novelty of the device, but also to the lack of transparency in their interactions with the virtual content.
Reflecting on how they appeared to bystanders, P4 remarked, ``I could picture myself just sitting there looking into blank nothingness and someone thinking, `What are you doing?' Like that looks so bizarre.''
Two participants
\textbf{anticipated that the device would become more socially acceptable over time} as people grew accustomed to its use.
P4, for instance, expressed that they would be more open to using AR if ``more people are familiarized'' with it.

In addition to issues with social acceptability, 
four participants 
reported that the device \textbf{hindered their ability to interact with others}.
This mostly stemmed from the glasses and virtual contents obstructing their field of view.
P7 and P8 reported having to look either above or below the lens to make direct eye contact. 
P4 removed his glasses entirely when talking with others, explaining, 
``I didn't want them to think ... that I'm not paying attention to them, ... that I'm just distracted by something else,'' such as a virtual window.

Participants' concerns about social acceptability likely contributed to their decision to complete their sessions primarily in private environments~($n$~=~111,~$N$~=~14).
In addition, 
several participants ($N$~=~4)
reported \textbf{adjusting their workspace arrangements around people} within their immediate surroundings. 
As P5 mentioned, ``If there were some people in front of me, or I was, like, talking to them while interacting with the interface, I tried to move [windows] away. I did not overlap my [windows] with their faces.''
P7 similarly preferred placing windows off-center when working among others, particularly strangers, to avoid staring at them while interacting with their own interfaces.
While P2 initially adopted a similar approach, ``mak[ing] an active effort to move screens out of people's faces,'' they later placed windows directly on top of people, reasoning that they could simultaneously view the virtual information while still maintaining a somewhat clear view of their conversation partner.

\subsection{Physical Comfort}
\label{sec:ergonomics}
Our results
indicate that many participants~($n$~=~17,~$N$~=~7) leveraged the device's flexibility in arranging windows to \textbf{adapt their workspace to their posture}, rather than adjusting their posture to fit the workspace.
As P1 remarks, ``I'm always hunched over my laptop so I have quite a bit of neck pain. But with this device, I didn't need to.''
P13 similarly adjusted their position while working and ``slightly shifted the window to match that.''

This flexibility also \textbf{allowed for positions that were previously unsupported by conventional laptops and computers}.
While most participants conducted the majority of their usage sessions at a desk or table~($n$~=~105,~$N$~=~14),
some explored engaging with their devices while resting in bed~($n$~=~16,~$N$~=~6) 
or reclining on their chair~($n$~=~5,~$N$~=~3).
P8 shared: ``We have those chairs that leaned back. I kind of put my whole canvas on the ceiling so I could lay back.''
P14 also expressed appreciation for being ``able to have a screen wherever I want, anywhere I want, anytime I want.'' 
They explained, ``Basically, I can be laying in bed eating. I always have a big screen in front of me. ''

However, considerations of physical comfort, to some extent, also limited usage of the AR workspace.
Several participants~($n$~=~8,~$N$~=~5)
reported that navigating the AR workspace to access distant elements strained their neck.
For example, when P4 first started using the device, they wanted to take advantage of the ``endless canvas'' and try out expansive workspace arrangements, such as ``stretch[ing] the screen up really high.''
However, 
they soon realized the physical toll, explaining, 
``I'm moving my head up, up and down, up and down, left and right. And I just started realizing, maybe that's the reason why my neck is hurting.''
This partially prompted several participants~($N$~=~7) 
to gradually move towards using fewer windows.
It is important to note that the aforementioned issues were likely exacerbated by the display's limited field of view.
As P2 notes, ``this field of view makes me move my head a lot to be able to see everything I need to.'' 

Similar considerations also contributed to the \textbf{preference for side-by-side layouts over vertical arrangements}~($N$~=~5).
In 9 sessions, 
participants arranged two windows vertically, one above the other.
Top-and-bottom layouts usually feature a primary task window and a secondary window that is only occasionally referenced. 
This contrasts with side-by-side layouts, which facilitate tasks that require more frequent cross-referencing.
Top-and-bottom layouts were used less frequently than side-by-side because participants found navigating horizontally more ``natural.'' 
As P5 remarked, ``most convenient was placing it side by side. I like that I don't have to turn my neck up to look for different screens.''

\subsection{Aesthetics}
\label{sec:aesthetics}

Several usage patterns we observed may be attributed to individual aesthetic preferences. 

\begin{figure}[t]
    \centering
    \includegraphics[width=0.9\linewidth]{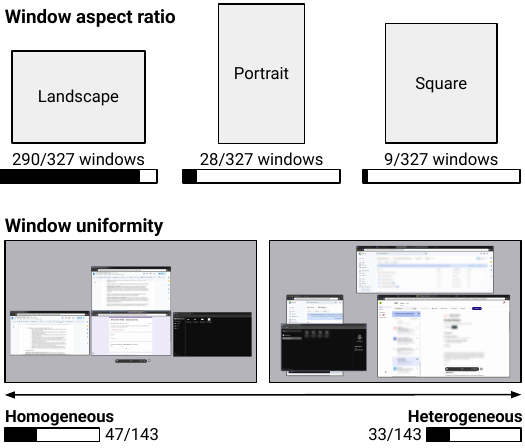}
    \caption{\textbf{Window sizing} --- Participants mostly sized their windows to have a landscape aspect ratio (290 out of 327 total windows across all sessions). In multi-window workspaces, window sizes ranged from homogeneous across all windows to heterogeneous.}
    \label{fig:window-sizing}
    \vspace{-1em}
\end{figure}

Participants \textbf{mostly sized their windows to have a landscape aspect ratio}~($n$~=~290,~$N$~=~14).
Portrait~($n$~=~28,~$N$~=~5)
and square~($n$~=~9,~$N$~=~6)
windows were used significantly less frequently in comparison (\autoref{fig:window-sizing},~top).
When participants had multiple windows in their workspaces, the scale and aspect ratios in these windows tended to be more uniform~($n$~=~47,~$N$~=~11), 
although a variety of display heterogeneity was represented nonetheless (\autoref{fig:window-sizing},~bottom).

\begin{figure}[t]
    \centering
    \includegraphics[width=0.9\linewidth]{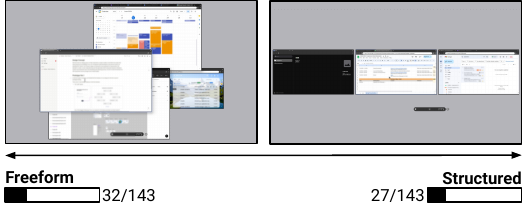}
    \caption{\textbf{Arrangement structure} --- Participants' workspaces varied from flexible arrangements to more structured grid layouts.}
    \label{fig:structure}
\end{figure}

Participants' multi-window workspaces \textbf{varied from a strict grid structure to more flexible arrangements} (\autoref{fig:structure}). 
Interestingly, participants generally preferred one of the two extremes.
In 32 sessions, 
participants' workspaces were largely unstructured, while in 27 sessions,
they were highly structured.

Several participants~($N$~=~6) \textbf{intentionally limited their window usage}.
As P10 noted, 
``I didn't want to feel overwhelmed with being surrounded by windows.''
In one session, P4 recalled, 
``I removed a screen off to the left side, not that it was distracting, but out of a sense of cleaning up my workspace.''

Some of these aesthetic decisions \textbf{may be the result of legacy bias}. 
As P12 remarked, ``I was still very biased towards what I'm familiar with. So, I usually use a two-device setup, or maybe a two-monitor setup, or a three-monitor setup. I find that I kind of copy that, in a way.''
Reflecting on their tendency to close unnecessary windows, P4 similarly commented, ``I think that may be a carryover habit from using a monitor, where you have a more finite amount of space.''


\subsection{Hybrid Usage}
\label{sec:hybrid}

\begin{figure}[t]
    \centering
    \includegraphics[width=0.9\linewidth]{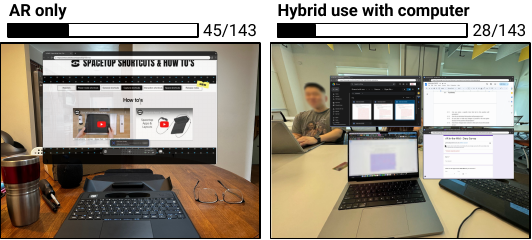}
    \caption{\textbf{Hybrid usage} --- Participants often used the AR laptop in combination with other devices, such as a laptop. \mr{The images shown are created based on participant descriptions, using the same compositing approach as in \autoref{fig:teaser},~\emph{right}.}}
    \label{fig:hybrid}
\end{figure}

\paragraph{\mr{Hybrid display}}
In 98 sessions, 
participants~($N$~=~14) reported \textbf{using the AR laptop in parallel with other physical displays}~(\autoref{fig:hybrid}), \mr{even though the device did not technically support cross-device interactions, as discussed earlier, due to the lack of tracking capabilities and software support for such interactions.
In such scenarios, participants manually positioned the virtual windows to fit their needs on the AR laptop, and switched inputs between devices opportunistically.}
These concurrent uses included phones~($n$~=~83,~$N$~=~14), 
computers~($n$~=~28,~$N$~=~9), 
tablets~($n$~=~1,~$N$~=~1), 
watches~($n$~=~4,~$N$~=~2), 
and TVs~($n$~=~2,~$N$~=~2).

Participants typically used a mobile phone to check and respond to notifications~($n$~=~34,~$N$~=~6),
messages~($n$~=~27,~$N$~=~9),
and emails~($n$~=~3,~$N$~=~2).
They also used their phone to write notes~($n$~=~3,~$N$~=~2), 
participate in calls~($n$~=~10,~$N$~=~6), 
and check the time~($n$~=~4,~$N$~=~4). 
Three participants favored a device with a physical display because it offered affordances that better accommodated their needs. 
For example, P2 found the ``slimmer screen'' of their phone to be better suited for social media scroll lists.
Three participants used a personal phone to separate their personal information and activities from their work device.
As P14 remarked, 
``I like to separate my everyday life from work-related things.''

When using the AR laptop alongside a conventional laptop or desktop computer, 
some users reported \mr{manually} arranging their virtual windows 
around a physical laptop~($N$~=~5).
\mr{
It is important to note that these arrangements required users to manually switch between the physical device's and the AR laptop's input.
}
For example, while P2 used their physical laptop for a design task, they displayed reference materials, such as a list of design requirements, as windows in AR and arranged them around their device.
P4 used a similar workflow, organizing reference materials around their computer monitor, which they used to author emails.
Two participants
used the AR laptop alongside physical devices due to the hardware limitations of the former, including its field of view and resolution.
They also relied on physical laptops to access applications and information that was unsupported or missing on the AR laptop ($N$~=~5).
Several examples of these applications include 3D graphics software~(P1) and code editors~(P13).

\paragraph{\mr{Hybrid task}}
Participants also \textbf{used the AR laptop simultaneously with physical tasks}~($n$~=~56,~$N$~=~11). 
For example, P3 used the AR laptop to display instructions that assisted in a prototyping task, aligning the window with their physical activities to engage in both tasks simultaneously.
P2 viewed videos in AR to entertain themselves while performing mundane tasks, like folding laundry.
P1 commented, ``I like having the freedom in my physical space, to multitask and not have a physical device getting in my way since it is a virtual workspace.''

\begin{figure*}[t]
    \centering
    \includegraphics[width=0.98\linewidth]{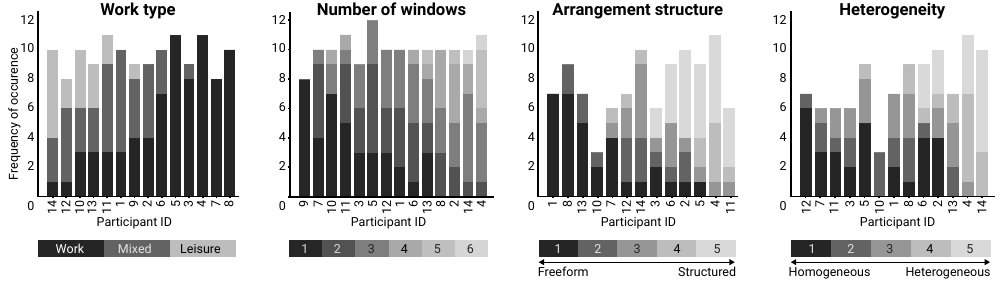}
    \caption{\textbf{Individual differences} - participants varied in the applications they opened, the number of windows they used, the amount of structure in their workspace arrangements, and the uniformity of their window sizes.}
    \label{fig:individual-diffs}
\end{figure*}

\subsection{Individual Differences}
In previous sections, we reported several behavioral tendencies that emerged from our aggregate observations. However, it is important to note that our participants also exhibited considerable variation in how they used the AR laptop.
As shown in \autoref{fig:individual-diffs}, participants had different applications in their virtual workspace, such as work, leisure, or a combination of both. They also varied in the number of windows used, the uniformity of window sizes, and the amount of structure in their workspace arrangements.
For example, while many participants ($N$~=~6) naturally opened an average of 3 to 4 windows, one participant exclusively used a single-window layout. 
Similarly, while participants ($N$~=~6) tended to configure their workspaces in a grid-like arrangement (\ie~4 or 5 arrangement structure rating on average),  others ($N$~=~5) followed almost no structure (\ie~1 or 2 arrangement structure rating).

\section{Values behind AR Laptop Usage}
In this section, we summarize the values that the participants gained from using the AR laptop. 

\paragraph{Physical Comfort}
As discussed in Section~\ref{sec:ergonomics}, the AR laptop offered flexibility in configuring virtual windows, enabling the creation of workspaces tailored to the user's posture. This flexibility notably supported postures that are not well accommodated by traditional devices such as laptops and desktops.

\paragraph{Portability}
Since the device's windows are virtual and occupy no physical real-estate, 
seven
participants also considered the device to be more portable.
As P13 noted, ``You can use this anywhere,'' in contrast to practices with current physical laptops and desktops that ``depend on specific setups.'' 
P8 envisioned that the device would enable working in ``smaller crowded spaces'' as a result.
P3 described an instance where they moved around with the device, walking between their living room and bedroom while maintaining continuity in their work. 

\paragraph{Multi-screen workflows}
Seven participants 
felt the device facilitated more efficient multi-screen workflows. 
It allowed windows to be flexibly configured into productive arrangements, such as side-by-side for cross-referencing~($n$~=~33).
The device's support for peripheral engagement was also regarded as beneficial~($N$~=~7).
Participants used their peripheral screen space to enable glanceable access to secondary content and to keep persistent windows for ongoing tasks.

\paragraph{In situ}
Six
participants appreciated the in situ nature of their windows. 
In particular, they believed that it allowed them to maintain a higher level of situational awareness of their physical surroundings~($N$~=~4).
For example, P11 regarded the ability to remain situated in and aware of the surroundings as critical for their role as an administrator: ``I would absolutely need it. If someone approached me, I would have to greet them.''
P12 noted that in a public setting, ``it is important to have a sense of what is happening in the surroundings,'' which is supported by the device.
As discussed in Sections \ref{sec:environment} and \ref{sec:hybrid}, 
the AR laptop also offered opportunities to supplement the user's task and physical environment with contextually relevant digital information.

\paragraph{Immersiveness}
Six participants 
reported feeling more immersed in their work while engaging with the AR device. 
P9 described the experience, ``the immersiveness was off the scale... I haven't experienced a flow state like that in a long time.''
P3 mentioned that while using the device, they felt ``less distracted from the real world in general.''
For P11, the immersive nature of the device also enriched leisure activities: ``Watching videos is a joy.... This could be the future of entertainment --- time flew by.''

\paragraph{Privacy}
Lastly, as discussed in Section~\ref{sec:social},
participants highlighted that the AR laptop offers potential privacy benefits.
They appreciated not ``hav[ing] to worry about [others] seeing information that isn't relevant to them,'' since all the AR laptop's contents were rendered through the head-mounted display (P9).

\section{Current Challenges with AR Laptop Usage}
In this section, we describe the challenges reported by our participants in using the AR laptop. 

\subsection{Hardware and software}
Participants encountered several hardware and software limitations with the AR laptop during their usage sessions. 
Seven participants found the \textbf{weight} of the glasses uncomfortable, 
while six participants reported issues with \textbf{overheating}.
Similarly, 
seven participants reported that the \textbf{display resolution} made it ``uncomfortable to visually look at'' (P6). 
Moreover, seven participants disliked how the glasses were tethered to their laptop keyboard, commenting that it ``limited [their] mobiliy'' (P7).
Finally, since the device only supported browser-based activities, it precluded some tasks that participants would like to have performed for their work~($N$~=~5).

Many of the aforementioned issues were effectively long-standing challenges within the AR research community~\cite{kim2018revisitingtrendsar}.
In our analysis, several limitations affected usage behaviors. 
For instance, as discussed in Section~\ref{sec:ergonomics}, the \textbf{restricted field of view} appeared to be one reason participants opted for more minimal setups, with several of them feeling limited ($N$~=~7).
For P10, this made using the device feel ``claustrophobic.'' 
P4 felt it restricted their interactions and task performance, noting, ``It was a little slower for me to complete my task.''
Some participants chose to avoid reading tasks due to visual discomfort caused by limited display resolution~($N$~=~5).
If their task involved reading, some participants chose to combine the AR laptop with higher resolution physical displays, instead of using it as a standalone device (Section~\ref{sec:hybrid}).

\subsection{Interaction limitations}
Hardware and software limitations aside, 
our results revealed several fundamental interaction limitations. 

\paragraph{\mr{Expanded display space navigation}}
While the expanded display provides benefits, such as improving multi-screen workflows, it also presents challenges for \textbf{navigation} across the display~($N$~=~3).
As P8 noted, ``the trackpad is too small for such a wide canvas.'' 
P2 similarly commented that they ``want to be able to make use of that entire space'' but disliked ``having to manage that using the mouse interface.''

\paragraph{\mr{Window management}}
The ability to flexibly add and configure windows also presents challenges for \textbf{window management}. 
Seven participants 
reported difficulties with having to manually arrange and align windows.
For example, P5 recalled, ``I did not like the range of the cursor movement on the virtual canvas. For instance, I was trying to enlarge the window but I had to click and drag it thrice to make it as big as I wanted it to be.''
P6 noted, ``I wish there was an automatic arrangement feature. [On my computer], when you want to put screens side by side, you can just drag [them into position] and it kind of partially [does] the task for you. But with this, I have to manually adjust every single detail.''
This suggests the need to introduce automated window management approaches, either through templates or more intelligent placement algorithms (\eg~\cite{cheng2021semanticadapt}). 

\paragraph{\mr{Discoverability}}
Three participants reported feeling \textbf{uncertain about how to best utilize the new medium}. 
As discussed in Section~\ref{sec:aesthetics}, participants tended to gravitate toward physical workspace arrangements they were already familiar with.
P13, for instance, ``did not really explore'' the different usage scenarios as a result. 
P14 commented, ``using this kind of device as a normal computer is not taking advantage [of] every single possibility that you can get with this kind of device.''
However, they found it challenging to conceive of new workflows, noting, 
``it's such a new concept that people are not used to thinking that way.''
To address this challenge, they suggested providing ``a guide on how to properly set it up, or a prompt saying, `you might want to do this like this.'{}''

\paragraph{\mr{Balancing virtual and physical}}
Lastly, 
it proved difficult for the device to effectively balance the \textbf{visibility} of the physical and virtual environments.
For instance, while several participants~($N$~=~6) found value in the immersiveness of the AR laptop, others reported feeling ``overwhelmed'' when too many windows were opened~(P10,~$N$~=~6).
Similarly, while participants appreciated the in situ nature of the displays~($N$~=~6), 
they also found their physical surroundings distracting~($N$~=~7). 
Combined, these results highlight several core design tensions in the development of AR systems.

%% file: sections/discussion.tex
\section{Discussion}
In this work, we investigated knowledge workers' usage and perceptions of AR for everyday tasks and activities.
Our findings confirm and extend many prior studies on AR. 
In this section, we relate these works to our results, discuss emergent usage patterns, identify opportunities for future research, and outline the limitations of our study.

\subsection{How Do People Work in AR?}
\label{sec:usage}

Our results indicate that participants adapted their use of AR based on several key factors: task, environment, social context, physical comfort, and aesthetics. 
These factors largely align with Grubert~\etal's~\cite{grubert2017pervasive} characterization of relevant context sources to consider in ``pervasive AR''.
We build on this work by highlighting how their considerations manifest themselves in the wild.

\paragraph{Virtual workspaces patterns}
In our study, participants engaged in tasks typical of knowledge workers (\eg~\cite{hernandez2024software}), including reading and writing documents, web browsing, and checking and composing emails. 
To support these tasks, participants explored various virtual workspaces.
Prior research has documented various patterns of window usage in AR, including the use of the central field of view for primary task content~\cite{pavanatto2024xrwild,lischke2016screenarrangement} and different window configurations for productivity tasks, such as vertical, horizontal, two-by-two, and personal cockpit layouts~\cite{mcgill2020seatedvrworkspace,ens2014personalcockpit}.
In our in-the-wild study, we observed participants using single windows and side-by-side layouts, which supported focused tasks and cross-referencing, respectively.
When more than three windows were used, 
they often represented applications for different tasks. 
Similar to the findings of Cho~\etal~\cite{cho2024minexr}, we observed participants blending leisure and work applications within their workspace in these settings. 
Additionally, consistent with Pavanatto~\etal~\cite{pavanatto2024xrwild} and Lischke~\etal~\cite{lischke2016screenarrangement}, 
for these multi-window layouts, 
participants organized their multi-window layouts around a central task area, while using the periphery for secondary applications.

Besides previously documented behaviors, we observed that participants used their peripheral screen space to maintain persistent AR windows. 
They sometimes stacked or overlapped their windows to support interactions such as quickly switching back and forth between window contents. 
They also adjusted window sizes according to their internal content, sometimes to form factors that are not supported by current screen-based computers.
Although much of previous work has focused on static layouts (\eg~\cite{cheng2021semanticadapt,biener2022vrweek}), 
our results indicate that virtual workspaces tend to be dynamically adjusted during usage. 

\paragraph{Environment}
Besides configuring their workspace based on task considerations, participants adapted window placements to account for environmental lighting conditions, 
as noted by Pavanatto~\etal~\cite{pavanatto2024xrwild}; 
geometric features and visual clutter, as highlighted by Cheng~\etal~\cite{cheng2021semanticadapt}; 
and the presence of others, as documented by Ng~\etal~\cite{ng2021passengerexperiencemrairplane} and Medeiros~\etal~\cite{medeiros2022shieldingar}. 
Participants also considered the perceived spaciousness of their environment in their workspace configurations, a factor that has not been previously documented. 

In addition, several participants attempted to use AR to support work from new postures and locations; however, consistent with McGill~\etal~\cite{mcgill2020seatedvrworkspace}, our results suggest that participants may require assistance in identifying optimal configurations.

\mr{
One interesting discrepancy we observed regarding the environment is that while some participants valued the device's support for more situated usage, others reported that the in situ display of virtual content obscured their physical surroundings. 
We believe this apparent contradiction stems from the inherent trade-off in AR between presenting virtual information and maintaining unobstructed real-world awareness.
Presenting information in situ can support a more seamless integration of virtual and physical tasks; however, introducing additional visual information will inevitably block some of a user's view of their physical environment. 
Depending on the scenario, this obstruction can be more or less acceptable. 
For example, if a user is passively working within a physical space, they may find that having virtual windows integrated into their environment offers an appropriate level of situational awareness, allowing them to feel present while attending to virtual tasks. 
However, if they are instead participating in a collaborative discussion, 
the visual occlusion can become a more significant impediment to their situational awareness and social interaction.
}

\paragraph{Aesthetics \& preferences}
Some of our findings on aesthetic considerations echoed prior work, particularly the tendency of some participants to arrange their workspace in a strict grid structure, consistent with the findings of Cheng~\etal~\cite{cheng2021semanticadapt}. 
However, this behavior was not representative of all participants, as some adopted more flexible arrangements instead, highlighting individual differences in preferences. 
Due to a combination of aesthetic preferences, device limitations, and legacy bias, many participants reported adopting more minimal setups, consisting of fewer windows that resembled their physical workspaces. This tendency contrasts with the multi-window layouts described in works such as Cheng \etal~\cite{cheng2021semanticadapt} and Lindlbauer \etal~\cite{lindlbauer2019contextaware}.
We believe these findings beneficially expand on previous understandings of how people may use window-based AR.

\paragraph{Hybrid usage}
While many of our participants used the AR laptop as intended, \ie~as a standalone laptop, 
we observed several participants adopting hybrid forms of usage. 
As Pavanatto~\etal~\cite{pavanatto2021virtualmonitor} previously showed, combining physical and virtual monitors may offer a better balance of familiarity with physical monitors, while also providing extra space from virtual monitors.
Our results showed that such a hybrid usage approach emerged in the wild when using an AR laptop, deriving similar benefits.
This highlights the potential for supporting integrated multi-device workflows with AR, rather than solely suggesting that AR should completely replace current devices, as previously proposed in Auda~\etal~\cite{auda2023crossreality}.
In addition to complementing physical screens, we also observed participants using the AR device in combination with physical tasks. 
This similarly echoes prior research suggesting that the unique value of AR lies in its seamless blending of virtual and real (\eg~\cite{lindlbauer2019contextaware}).

\mr{
\paragraph{AR versus physical displays}
Our study highlights several advantages of AR over physical displays in the wild, including improved physical comfort, portability, multiscreen workflows, in situ display, immersiveness, and privacy. 
These advantages are largely enabled by AR's affordances, which support unique usage patterns. 
One notable pattern facilitated by AR's expanded display is the utilization of peripheral spaces to host persistent windows and support multiple parallel tasks, a capability previously limited by the finite screen real estate of physical displays.
Similarly, AR's flexibility in workspace arrangements, a contrast to the fixed nature of physical screens, enabled users to design window layouts for new usage postures, such as reclining in bed.
Last but not least, by enabling the in situ display of virtual content, we observed AR supporting hybrid workflows that allowed users to simultaneously perform physical tasks and access virtual information. 
Unlike physical displays, which demanded either the user's hands or dedicated space, AR's less intrusive, hands-free display approach provided more freedom in accomplishing diverse physical activities.
   
In our study, we also observed notable similarities in usage patterns. 
For instance, many workspace arrangements mirrored common screen-based window layouts.
Similarly, while the device had an expanded display space, our results also indicate that users commonly designed workspaces that contained a small number of windows (\ie~2.4 windows on average), which could also have been supported by physical display setups.
We speculate that this may be due to legacy biases, system limitations, and the nature of the sort of tasks our users performed. 
First, the AR laptop is largely designed as a display extension rather than a fundamentally different interface. 
This may have prompted users to approach it with a mental model similar to their existing devices. 
Second, the AR laptop's hardware limitations, notably its limited field of view, could have restricted the range of workspace usage. 
Furthermore, because the AR laptop exclusively supported 2D browser-based applications, users could not fully explore more AR-specific use cases. 
Finally, for the everyday knowledge-work oriented usage scenarios, it is possible that users did not inherently need to leverage AR's unique affordances, such as support for a greater number of virtual windows.

Overall, 
while the AR laptop enables several new and unique workflows, 
users also adopted usage patterns that sometimes mirrored those established with physical displays.
}


\subsection{Opportunities}
Our study offers a glimpse of what prolonged working in AR may look like in the future, including the advantages it offers and the challenges it presents. 
Our results highlight several opportunities for future work. 

First, our study shows that current devices make it possible to use AR for everyday tasks outside of controlled laboratory environments. In certain situations, regular users can experience substantial benefits. Although there are still hardware and software barriers, we believe this indicates that AR will become increasingly integrated into daily activities in the near future.
As such, it is pertinent for our research community to continue rigorously and longitudinally examining how users might interact with AR in their daily lives, both for knowledge work and beyond.
We expect improvements in AR devices to address technical issues such as limited field of view and low resolution, potentially leading to changes in user behavior.
For example, increasing the field of view of AR devices might allow users to take better advantage of expanded display space.
However, technical developments will take time, and even so, imperfections are likely to persist.
Therefore, examining user behavior with current devices and tools is invaluable.

Our results also indicate that in addition to tackling hardware and software challenges, \textbf{several fundamental interaction issues must be addressed} to facilitate the effective daily use of AR. 
One common issue we observed was efficiently \textbf{navigating the expanded AR display space}.
\mr{
Efforts within the current literature have explored ways to support more efficient navigation, including the use of gaze as an additional input modality~\cite{biener2020breakingscreen}.
}
We encourage researchers to build on these efforts, particularly by studying ways to improve this process outside controlled laboratory settings.
Another challenge relates to \textbf{window management}. 
To address this issue, we believe there is value in operationalizing common arrangement patterns we observed as layout templates that users can snap windows to, similar to the application tiling feature found in macOS~\cite{macOStiling}.
However, given the sensitivity of window arrangements to their surroundings, more intelligent adaptation approaches may be necessary. 
\mr{
Several prior works have explored optimization-based approaches to adapting interfaces based on various contextual factors, such as environment semantics~\cite{cheng2021semanticadapt}, user placement preferences~\cite{niyazov2023layoutoptimization}, and usability requirements~\cite{belo2022auit}. 
Building on Lu and Xu's findings~\cite{lu2022spatialuitransitions}, we speculate that future interfaces will need to support a combination of manual arrangement and automatic adaptation. We encourage future work to continue investigating such mixed-initiative approaches.
}

Despite improvements in hardware, software, and interactions, challenges may still persist in promoting adoption and effective usage. 
As discussed in Section~\ref{sec:usage}, we found that users tended to revert to their usual window usage habits in physical workspaces.
In many cases, these usage patterns did not take advantage of the unique affordances of AR. 
Participants themselves even reported feeling unsure about how to effectively use the new medium.
A possible solution to these challenges could involve \textbf{improving upon the onboarding process}. Although we attempted to thoroughly explain the device functionalities during our introductory session, providing more detailed and repeated training might be helpful.
To \textbf{help users overcome their legacy biases} regarding window arrangement patterns in particular, future work could consider exploring context-based and situational onboarding methods~\cite{chauvergne2023onboarding}, such as suggesting layouts tailored to their current task.
Yet, while we see AR as a means to engage with digital information in new ways, 
we also recognize the importance of maintaining familiar interfaces for work to facilitate a smoother transition from current laptops and devices.
Future devices \textbf{should leverage familiar aspects while introducing new features}, but this is likely easier said than done, requiring careful design of onboarding procedures and interfaces.
Perhaps parallels can be drawn to the introduction of mobile user interfaces. 
Early mobile user interface designs heavily relied on the preceding desktop paradigm~\cite{punchoojit2017mobileui}. While these designs did not initially suit the target form factor, 
they nonetheless provided a familiar entry point for users.
AR interfaces today may be in a similar state.
As previously discussed by Pavanatto~\etal~\cite{pavanatto2024multiplemonitors}, 
drawing on familiar paradigms may ease the transition from today's personal computers to computing with virtual displays, thereby enabling users to adapt more quickly and effectively to the new medium.

Last but not least, 
with AR entering our daily lives, 
as previously suggested by Abowd and Mynatt~\cite{abowd2000chartingubicomp},
how it will be used will differ significantly from controlled laboratory settings. 
In particular, people's activities will often lack clear beginnings or ends, sometimes occurring at the same time.
In our study, we observed patterns such as the integration of productive and leisurely tasks. AR devices designed for productivity must consider how people actually work, including providing opportunities to disengage~\cite{kaur2020breaks}.

\subsection{Limitations}
Our current study design is subject to several limitations.
First, while we aimed to study usage behaviors longitudinally, 
our results currently reflect the participants' engagement with the device for only 40 minutes a day over a two-week period.
\mr{It should also be noted that participants received a monetary incentive to complete their sessions, which may have introduced biases in how and when they chose to use the device.}
To capture behaviors that may emerge if the device completely replaces users' current laptops or computers, 
we encourage researchers to continue studying AR usage over longer periods. 
However, it is also 
it is important to acknowledge the significant logistical and feasibility challenges involved. 
An alternative approach to consider is to consult and gather feedback from current power users.

Additionally, in our work, 
we based our analysis on post-usage diary surveys and photographs, as well as interviews.
Surveys and interviews provided valuable qualitative insights into participants' usage; however, they might be affected by memory biases since they required participants to recall their experiences.
Our photographs also provide rich information, providing a visual context for participants' written reflections.
However, while our results showed that participants often dynamically adjusted their virtual workspaces during their sessions, the photographs only capture the final state of these workspaces.
Our study would have benefited from collecting continuous telemetry or video data of participants' usage sessions; however, we chose not to out of concern for their privacy.
That said, we still believe that collecting more granular data will be valuable in the future, but such measures should be carefully weighed against their invasiveness.

It is also important to acknowledge that, while our qualitative data provides detailed insights into how a small group of participants uses and perceives AR laptops, it does not allow for statistical evaluation of the prevalence of these behaviors in a larger population. 
To validate these results, further quantitative studies would be beneficial. 
For instance, 
building on our observations of potential individual differences in AR laptop usage, it would be valuable to examine a larger population to determine whether users fall into distinct behavioral categories (\eg~preference for freeform or structured arrangements). 
Additionally, studying a population beyond a knowledge worker or university context would help assess the generalizability of our results.

Furthermore, 
our study was based on a specific hardware and software implementation of AR, which contains known limitations and is not wholly representative of future AR devices. 
It is not unlikely that as AR devices improve, user behaviors and perceptions may also change. 
For example, if the device supported more applications and was better integrated with participants' individual information ecosystems, they might have used the device more exclusively, rather than in combination with other computers.
\mr{
Similarly, with an expanded field of view, participants may be more inclined toward using the peripheral spaces within their display. 
With greater adoption, we may also expect social norms to change, encouraging more public usage of the device and a distancing from legacy biases and current aesthetic preferences. However, we believe many of our findings will also persist.
For example, we expect that comfort will remain a key consideration in workspace design. Specifically, we speculate that certain window arrangements, such as single-window or side-by-side usage, will probably remain in use, particularly in contrast to top-and-bottom arrangements, because they facilitate more ergonomic usage. 
Additionally, if users continue to employ AR laptops for tasks similar to those observed in this study, we speculate that their workspace arrangement strategies will also remain consistent. For instance, for the same tasks, users will likely not change the number of windows used or how they are arranged, as the task structure will likely necessitate the same interactions.
Replications of our study with future devices would yield valuable insights into how usage patterns may evolve or persist.
}

Finally, our findings only capture current user perceptions and use of AR, situated within a technological landscape still dominated by mobile phones and computers. 
At the time of writing, AR was primarily used by early adopters only.
As AR usage becomes more prevalent, we may expect people's attitudes toward the technology to change. 
For example, although our participants expressed reluctance to use the device in public, social attitudes toward the device are likely to change as AR usage becomes more commonplace.
This suggests a need to continuously re-evaluate how people may use AR going forward.

%% file: sections/conclusion.tex
\section{Conclusion}
In this work, we contribute insights from a longitudinal diary study with 14 knowledge workers, who used an AR laptop for around 40 minutes a day over two working weeks.
We collected reflections on over 104 hours of work (7 hours per participant) through diary surveys, photographs, and interviews. 
During their usage sessions, participants engaged in a variety of tasks across different physical environments. 
Our findings shed light on how task, environment, social, and aesthetic factors influenced participants' usage patterns. 
Participants' tasks determined the number, arrangement, and size of virtual windows they used. 
External factors such as lighting, visual clutter, physical constraints, and the presence of others guided participants' window placements.
Participants further arranged their workspace in AR to facilitate more comfortable workflows, 
while making certain design decisions based on personal aesthetic preferences. 
Some participants used the AR laptop in combination with physical displays and tasks. 
Finally, we document the values and current challenges associated with using AR laptops, as reported by our participants.
Our results provide insight into how knowledge workers might use AR in the future, presenting numerous opportunities for further exploration. 
We believe it is crucial for the research community to continue developing an ecologically valid understanding of how AR may be used, especially in longitudinal studies, in preparation for its future pervasive use.

%% file: sections/acknowledgements.tex
\acknowledgments{
We thank our project sponsors and collaborators from Sightful, particularly Oren Zuckerman, Tom Hitron and Tamir Berliner.
This research was partially supported by Sightful, which provided funding and hardware. 
The authors gratefully acknowledge their contribution to this project. 
We also thank all involved peers and participants. 
Yi Fei Cheng received partial support from the Croucher Foundation. 
The views expressed in this paper are those of the authors and do not necessarily reflect the views of Sightful.
}

%% file: sections/biography.tex
\begin{center}
\begin{tabular}{p{1in} p{2.3in}} 
     \vspace*{-10px}
    \includegraphics[width=1in,height=1in]{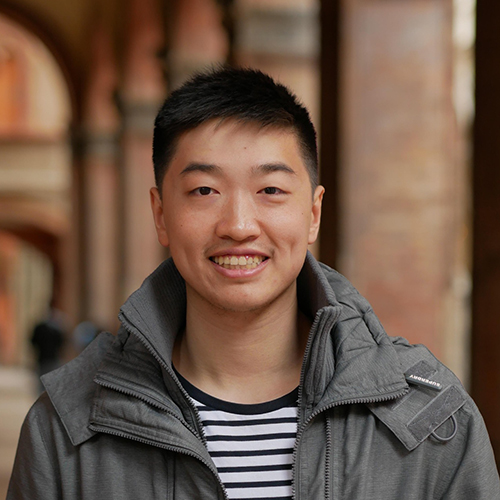}
    &
    \begin{minipage}[t]{2.3in}
        \textbf{Yi Fei Cheng.} \\
        Yi Fei Cheng is a Ph.D. student in Human-Computer Interaction at Carnegie Mellon University.
        He is a member of the Augmented Perception Lab.
        His research interests include adaptive interfaces and extended reality.
    \end{minipage} \\ 
    \vspace{10pt} \\ 

    \vspace*{-10px}
    \includegraphics[width=1in,height=1in]{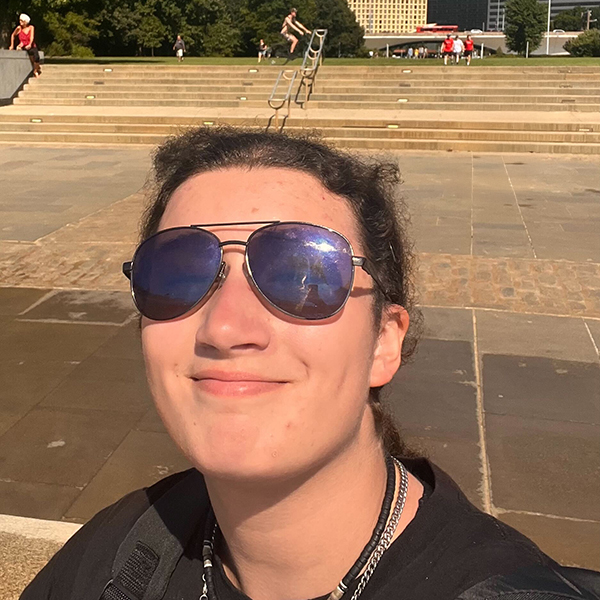}
    &
    \begin{minipage}[t]{2.3in}
        \textbf{Ari Carden.} \\
        Ari Carden is a research staff at the Augmented Perception Lab at Carnegie Mellon University.
        Her research focuses on extended reality for productivity.
    \end{minipage} \\ 
    \vspace{10pt} \\ 

    \vspace*{-10px}
    \includegraphics[width=1in,height=1in]{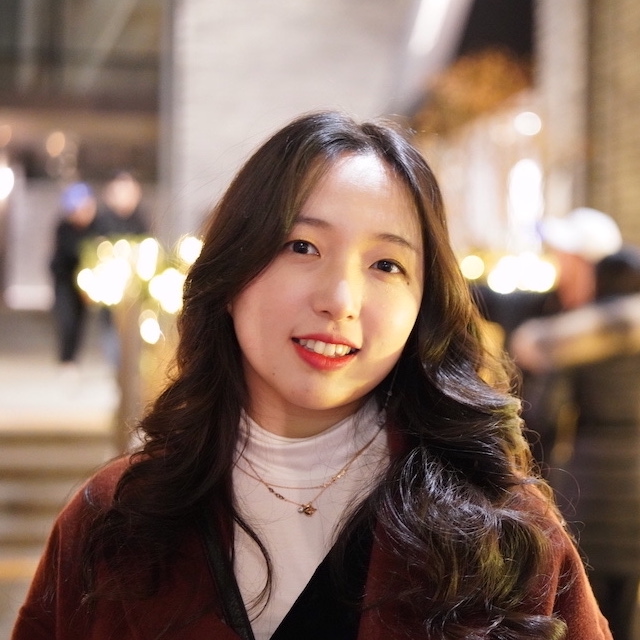}
    &
    \begin{minipage}[t]{2.3in}
        \textbf{Hyunsung Cho.} \\
        Hyunsung Cho is a Ph.D. student in Human-Computer Interaction at Carnegie Mellon University.
        She is a member of the Augmented Perception Lab.
        Her research focuses on developing perceptually intelligent user interfaces for extended reality.
    \end{minipage} \\ 
    \vspace{10pt} \\ 

    \vspace*{-10px}
    \includegraphics[width=1in,height=1in]{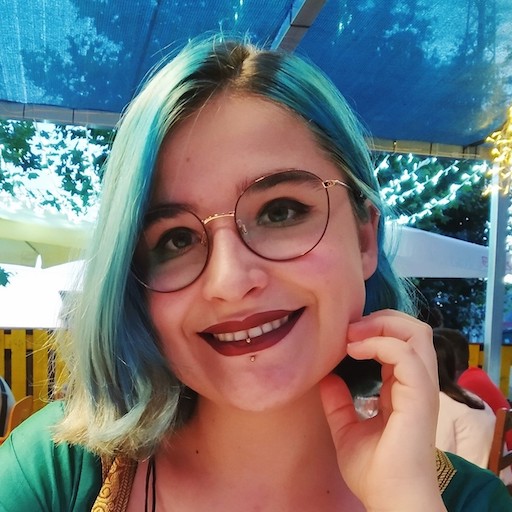}
    &
    \begin{minipage}[t]{2.3in}
        \textbf{Catarina G. Fidalgo.} \\
        Catarina G. Fidalgo is a Ph.D. student in Human-Computer Interaction at Carnegie Mellon University.
        She is a member of the Augmented Perception Lab.
        Her research interests include remote collaboration and extended reality.
    \end{minipage} \\ 
    \vspace{10pt} \\ 

    \vspace*{-10px}
    \includegraphics[width=1in,height=1in]{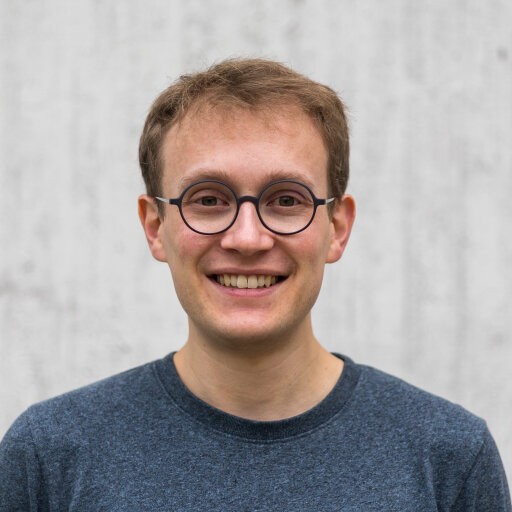}
    &
    \begin{minipage}[t]{2.3in}
        \textbf{Jonathan Wieland.} \\
        Jonathan Wieland is a Ph.D. candidate in the Human-Computer Interaction Group at the University of Konstanz. His research interests include exploring the potential and challenges of extended reality-supported interactive exhibitions.
    \end{minipage} \\ 
    \vspace{10pt} \\ 

    \vspace*{-10px}
    \includegraphics[width=1in,height=1in]{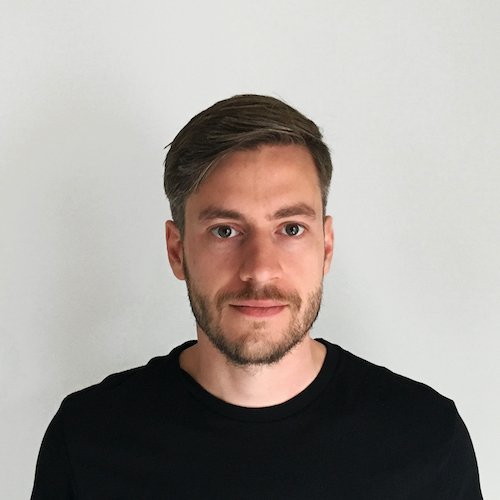}
    &
    \begin{minipage}[t]{2.3in}
        \textbf{David Lindlbauer.} \\
        David Lindlbauer is an Assistant Professor at the Human-Computer Interaction Institute at Carnegie Mellon University, leading the Augmented Perception Lab. 
    \end{minipage} \\ 

\end{tabular}
\end{center}

%% file: sections/supplementary.tex
\clearpage

\setcounter{page}{1}

\renewcommand{\thefigure}{A\arabic{figure}}
\setcounter{figure}{0}     

\renewcommand{\thetable}{A\arabic{table}}
\setcounter{table}{0}   

\renewcommand{\thesection}{A\arabic{section}}
\setcounter{section}{0}   

\begin{center} 
\textbf{SUPPLEMENTARY MATERIAL} 
\\[1em]
\centering
Augmented Reality Productivity In-the-Wild: \\
A Diary Study of Usage Patterns and Experiences of Working with AR Laptops in Real-World Settings
\end{center}

\mr{
\section{System Implementation and Interaction Capabilities}
In the following, we document the technical specifications and interaction capabilities of our deployed Spacetop EA (Early Access) devices. 
The Spacetop EA (Early Access) device is an early beta version of the commercially available Spacetop Augmented Reality solution.
It was designed as an all-in-one AR laptop and used for external beta testing.
While the EA version was a standalone unit, the latest version of Spacetop is compatible with commercial laptops.
\subsection{Technical Specification}
\begin{table}[h]
    \centering
    \begin{tabular}{p{0.25\linewidth} p{0.6\linewidth}}
        \toprule
        \textbf{Model} & SPEA001 \\
        \midrule
        \textbf{Chip} & Qualcomm Snapdragon XR2, Kryo 585TM 8-core 64-bit CPU, up to 3.1 GHz speed, AdrenoTM 650 GPU \\
        \midrule
        \textbf{Battery life} & 5+ hours \\
        \midrule
        \textbf{Memory} & 8GB LPDDR5 \\
        \midrule
        \textbf{Charging and expansion} &  Charger: 65W USB-C, 110v/220v, 50Hz/ 60Hz, US plug 2x USB-C sockets, supporting: PD3.0 fastcharging(up to 65W), Charge from 0\% to 85\% in less than 2 hours, SuperSpeed USB up to 10Gb/s \\
        \midrule
        \textbf{Connectivity} & Wi-Fi 6 802.11ax, Bluetooth 5.1, Nano-SIM card (SIM included), 5G/LTE NR Sub-6 \\
        \midrule
        \textbf{Webcam} & 5 MP, Resolution 2560 x 1920 \\
        \midrule
        \textbf{Keyboard and touchpad} & Full-sized keyboard and clickable touchpad \\
        \midrule
        \textbf{Mini-Display} & 1.54'' e-paper display, Black and white,  200x200 pixels display resolution \\
        \midrule
        \textbf{Operating system} & Spacetop OS -- Custom flavor AOSP \\
        \midrule
        \textbf{AR Glasses: Xreal Light} & Display \& Audio: 2 OLED display panels (1920x1080 pixels per eye) with sRGB 106\% color gamut, 72Hz refresh rate, 8-bit depth for 16.773 million colors, up to 280 nits perceived brightness, 100,00:1 contrast ratio, \~52 degree FoV (Diagonal), 42 pixels per degree (PPD), fingerprint-resistant coating, 2 open-ear speakers, dual microphone array, omnidirectional MEMS, microphones \\
        \midrule 
        \textbf{Screen size} & Equivalent to 100 inches \\
        \midrule
        \textbf{Audio} & 3.5 mm headphone jack
        \\
        \midrule
        \textbf{Dimensions} &  Height: 1.57 inches (4 cm), Width: 10.47 inches (26.6 cm), Depth: 9.8 inches (24.9 cm) \\
        \midrule
        \textbf{Operating temperature} & $50^{\circ}$ to $95^{\circ}$ F ($10^{\circ}$ to $35^{\circ}$ C) \\
        \midrule
        \textbf{Storage temperature} & $14^{\circ}$ to $113^{\circ}$ F ($-10^{\circ}$ to $45^{\circ}$ C) \\ 
        \bottomrule
    \end{tabular}
    \caption{Spacetop EA Technical Specifications}
    \label{tab:techspecs}
\end{table}

\subsection{Interaction Capabilities}
The Spacetop EA device enables users to open, resize, move, and interact with windowed browser-based applications on top of a virtual "spatial canvas." 

\subsubsection{The Spatial Canvas}
The spatial canvas is a virtual plane where users organize their workspace windows. It is the first virtual interface element shown on start-up, demarcated by grid-like dots.
The canvas's size is 100 inches diagonally. 
By default, it is world-anchored~\cite{feiner1993windows} and curves over a cylinder surrounding the user, similar to the concept described by Pavanatto~\etal~\cite{pavanatto2024multiplemonitors}.
The canvas supports the following interactions:
\paragraph{Zooming in/out}
Users can zoom in or out to adjust the spatial canvas's distance, which effectively changes the apparent size of its contents. Zooming in brings the canvas closer, making windows appear larger within the field of view. Zooming out moves the canvas further away, shrinking the content and allowing for greater visibility of windows across the overall canvas.
Users can zoom in or out by performing a three-finger drag gesture on the trackpad. Dragging down zooms in, and dragging up zooms out.

\paragraph{Panning}
Users can shift the canvas left or right by performing a three-finger drag gesture in the desired direction on the trackpad.

\paragraph{Adjusting height}
Users can shift the canvas vertically. 
To adjust its height, they must hold Control, the Spacetop key, and Alt, then press Up or Down accordingly.

\paragraph{Tilting}
Users can change the canvas's angle by holding Control while performing a three-finger drag up or down on the trackpad.

\paragraph{Re-centering}
Users can recenter their canvas to their current facing direction by pressing both the left and right Shift keys.

\paragraph{Changing anchoring mode}
By default, the device anchors the canvas to the world. 
Users can toggle between this default setting and a "travel mode," which anchors the canvas to the device's keyboard instead.
Pressing the Spacetop key and "T" together toggles between these two modes.

\subsubsection{The Home Bar}
A "home bar" menu is persistently displayed at the base of the spatial canvas. 
It hosts core system functionalities such as shutting down, sleeping, or restarting the system, and provides access to settings. 
The home bar also offers quick access to various system features, including Wi-Fi and Bluetooth connectivity, live streaming, a photo app, a video recording app, volume controls, and charge status. 
Finally, it contains a launcher for opening and accessing applications.

\subsubsection{Virtual Windows}
Users can open applications as virtual windows on top of the spatial canvas. 
Windows can be moved around and resized using direct manipulation. 
Specifically, users can move windows by clicking and dragging their header bar, a gray linear strip located at the top of every window.
Users can resize windows by clicking and dragging their corners in the desired direction.

\subsubsection{Physical Contextual Awareness}
Beyond persistent world-anchoring of the spatial canvas, at the time of the study, the device did not support other physical contextual awareness mechanisms, such as knowledge of surfaces or physical devices for virtual content placement.
}

\section{Background Questionnaire}
Our background questionnaire included the following items. 
\begin{enumerate}
\item[Q1] How old are you?
\item[Q2] How would you describe your gender?
\item[Q3] What is your current job title?
\item[Q4] What is the highest level of education you have completed?
\item[Q5] Please describe the devices you currently own, as we show below:

Device Number - Type of Device - Device Brand - What you use the device for (multiple uses OK) - Approximate usage frequency

For example:
\begin{enumerate}
\item[1] Phone - Google Pixel 4 - For personal usage - Every day for about 2 hours

\item[2] Phone - Apple iPhone 11 - For work usage - Every day for about 2 hours

\item[3] Tablet - Microsoft Surface Pro 6 - My main work device - Every day for about 5-6 hours

\item[4] Tablet - Kindle Fire - Tablet for kids - Infrequently, only to troubleshoot for the children

\item[5] Smart Watch - Apple Watch - Exercise tracking, answering phone calls - Almost always on me, though checking infrequently
\end{enumerate}

\item[Q8] On a scale of 1 to 7, rate your previous experience with Augmented Reality (\eg~Microsoft HoloLens, Google Glass)

\item[Q9] Describe your previous experience with Augmented Reality. 
\item[Q10] On a scale of 1 to 7, rate your previous experience with Virtual Reality (\eg~Oculus Quest, Valve Index)

\item[Q11] Describe your previous experience with Virtual Reality. 

\item[Q12] Do you have any known visual impairments? If so, please indicate the type of visual impairment (e.g., color blindness). 
\item[Q13] Do you wear contact lenses?
\item[Q14] Do you wear glasses?
\end{enumerate}


\section{Onboarding Protocol}

\textbf{<0 Introduction> (20 minutes)}

\begin{enumerate}[label=(\arabic*)]
\item Hello, my name is []. I am a researcher at the [] lab. 

\item This study is part of a project aimed at understanding how people use AR for productive work. The primary objective of this study is to gather feedback on your experiences with an AR device in your everyday work environments.

\item For the study, upon obtaining your consent to participate, we will first ask you to complete a demographics survey, which includes questions about your background, device usage habits, and experiences with AR and VR. 

\item We will then introduce you to the AR device. 

\item Afterwards, we will ask you to use the device for at least 10 sessions over the upcoming two weeks, one session per day, with each session lasting at least 30 minutes. 

\item Following each session, we request that you document your experience in a diary survey.
Note that in total, this amounts to around a 40-45 minute commitment per day. 

\item Finally, you will be asked to participate in a 30-45 minute semi-structured interview to reflect on your overall experience.

\item Please note that your participation in this study is entirely voluntary, and you can interrupt your involvement at any time without the need for justification. 

\item In the course of the study, we will collect your survey responses, as well as some photographs of your virtual workspace and environment. Towards the conclusion of the study, we will also record our interview with you. 

\item These interview recordings will be transcribed and, together with the survey data, will be analyzed by our research team. Our plan is to publish our findings cumulatively and anonymously.

\item Do you have any questions at this point?
\begin{enumerate}
\item Answer questions.
\item Administer consent form.
\item Administer background questionnaire.
\end{enumerate}
\end{enumerate}

\textbf{<1 Device Setup and Instructions> (30 minutes)}

\begin{enumerate}[label=(\arabic*)]
\item For our study, you will be using the Sightful Spacetop EA device. Before I go into the detailed study tasks, let me introduce the device.

\item The Spacetop device is effectively a laptop without a screen. Instead of a screen, its virtual contents are displayed via a pair of augmented reality glasses. 

\begin{enumerate}
\item Demonstrate powering on and off the device.
\item Demonstrate basic system functions, 
such as connecting to networks, 
adjusting display brightness,
changing device settings, capturing screenshots, 
and accessing applications through the dock.
\item Demonstrate interactions for window management, including opening, closing, positioning, centering, and resizing of windows.
\end{enumerate}

\item Participants were allotted supervised time to familiarize themselves with the device, receiving assistance from the experimenter as needed. The study proceeded to the next stage only after participants confirmed their readiness.
\end{enumerate}

\textbf{<2 Study Tasks Explanation> (10 minutes)}

\begin{enumerate}[label=(\arabic*)]
\item Now, let us go into the study tasks in more detail. 

\item Over the next two weeks, please use the device at least once a day for a minimum of 30 minutes, for a total of 10 days. If you miss more than two days in a row, we will, unfortunately, have to discontinue the study.

\item You are encouraged to complete additional sessions beyond the minimum. We will pay you \$15 for each extra session.

\item Feel free to use the device at any time or place and for whatever task you choose.

\item After each session, we would like you to complete a diary survey, reflecting on your usage experience. 
You will receive a link to the survey via email shortly after the onboarding session. 
We have also written them on the device for your convenience. Please try to complete the survey immediately after using the device, or as soon as possible afterward.

\item Do you have any questions at this point?
\begin{enumerate}
\item Answer questions.
\end{enumerate}

\item Lastly, if you have any questions or trouble with the device at any point, contact us immediately. 

\end{enumerate}

\section{Diary Survey}
Our diary survey included the following items. 
\begin{enumerate}
\item[Q1] What was the approximate start time of your session?
\item[Q2] What was the approximate end time of your session?
\item[Q3] What specific tasks did you engage in the AR Laptop?
In a bulleted list, describe the tasks you have undertaken. 

For example:
\begin{itemize}
    \item[-] Read and annotated an assigned text on a PDF reader
    \item[-] Searched for the definition words using a web browser
\end{itemize}

\item[Q4] What other tasks or activities did you engage in outside of the AR Laptop (e.g., in real life, on other devices)?
In a bulleted list, describe the tasks or activities you have undertaken. 

For example:
\begin{itemize} 
\item[-] Responded to notifications on my mobile device
\item[-] Briefly conversed with a colleague seated across from me
\item[-] Hand-wrote notes while reading news articles
\end{itemize}

\item[Q5] Where did you complete this session (i.e., location)?

\item[Q6] What were the physical surroundings in the location during your session? 
In bullet points, describe the physical surroundings and items in the location where you completed the session.

For example:
\begin{itemize}
\item[-] Spacious, open atrium with high ceilings
\item[-] On a shared working desk, with people chatting across
\item[-] Occasional sounds of coffee machine
\end{itemize}

\item[Q7] How did you construct your virtual workspace on the AR laptop? 
In 2-3 sentences, describe what applications you used and how they were arranged in space.

\item[Q8] If you changed your virtual workspace during the session, what adjustments did you make?
In 2-3 sentences, describe any modifications you made to your virtual environment, and explain your reasons for these changes. Examples of modifications might include adjustments to window size, changes in window layout, opening and closing applications, and modifying the depth of virtual elements, among others.

\item[Q9] Describe the factors that influenced how you constructed your virtual workspace (e.g., task, physical surroundings)?
In bullet points, describe the factors that influenced how you constructed your virtual workspace. Provide detailed examples for how each factor influenced the workspace design.

\item[Q10] What aspects of the virtual workspace did you like in this session?
Describe the aspects you liked in a bulleted list. Some directions to think about include but are not limited to the device's influence on your productivity and comfort.

\item[Q11] What aspects of the virtual workspace did you dislike in this session?
Describe the aspects you disliked in a bulleted list. Some directions to think about include but are not limited to the device's influence on your productivity and comfort.

\item[Q13] If possible, can you upload a screenshot of your virtual workspace?

\item[Q14] If possible, can you upload a photograph of your current physical environment? 
\end{enumerate}

\section{Interview Protocol}


\textbf{<0 Introduction> (3 minutes)}
\begin{enumerate}[label=(\arabic*)]
\item Hello. Thanks for completing the usage sessions and agreeing to partake in this final interview. 
\item In this interview, I will ask you to reflect on your experiences using the AR device. I would like to explore the tasks you performed, how you set up your virtual workspace, and any external factors that influenced your use of the device. Additionally, I will ask questions about how you integrated the AR device with other physical devices and tasks, as well as your overall perceptions of the device.

\item The interview will be recorded so that we can analyze it later. May I have your permission to begin recording?
\end{enumerate}

\textbf{<1 Tasks> (7 minutes)}

\begin{enumerate}[label=(\arabic*)]
\item 
What tasks did you perform with the AR device, and how did your usage change over the 10 sessions?

\item
Which tasks were notably well-suited or ill-suited for AR, and why? Were there tasks you intended to do but were restricted by the device?

\item
How did the experience of using the AR device for these tasks compare to regular screen-based computers?
\end{enumerate}

\textbf{<2 Workspace Arrangement> (10 minutes)}

\begin{enumerate}[label=(\arabic*)]
\item 
How did you customize your virtual workspace, including changes across sessions? How did task nature and physical surroundings influence your setup?

\item 
How did your workspace change within sessions and did this differ across time? 

\item 
How did your workspace arrangement affect your productivity and comfort? Were there limitations in customization due to the device?
\end{enumerate}

\textbf{<3 Integration with Physical Tasks> (5 minutes)}

\begin{enumerate}[label=(\arabic*)]
\item
Did you engage in non-AR activities while working in AR? Describe how you transitioned between them.

\item
Did you engage with other devices while working in AR? Describe how you transitioned between them.

\item
Can you envision a seamless integration of AR workspaces with daily activities and other devices?

\end{enumerate}

\textbf{<4 Perceptions> (10 minutes)}

\begin{enumerate}[label=(\arabic*)]

\item What were the most beneficial aspects of the AR workspace, and how did they impact your activities? Conversely, what limitations or disadvantages affected your experience?

\item Did your perception of the device's advantages or disadvantages change over time?

\item What improvements or features would enhance your AR experience? Based on your experience, how do you see the future of AR in daily tasks?
\end{enumerate}